\AtBeginDvi{}

\documentclass[reprint,groupedaddress]{revtex4-1}
\usepackage[dvips]{graphicx}
\usepackage{amsthm}
\usepackage{amsmath}
\usepackage{physics}
\usepackage{here}
\usepackage{amssymb}
\usepackage{color}
\usepackage{bm}% bold math
\usepackage{bbold}
\usepackage{letltxmacro}
\usepackage{algorithm}
\usepackage{algorithmic}
\usepackage{threeparttable}
\usepackage{caption}
\usepackage[usenames,dvipsnames]{xcolor}
\usepackage{url}
\usepackage[breaklinks,colorlinks=true,citecolor=blue,linkcolor=RubineRed,urlcolor=RubineRed]{hyperref}
\begin{document}
\title{SantaQlaus: A resource-efficient method to leverage quantum shot-noise for optimization of variational quantum algorithms}

\author{Kosuke Ito$^{1}$}\email{kosuke.ito.qiqb@osaka-u.ac.jp}
\author{Keisuke Fujii$^{1, 2, 3}$}\email{fujii@qc.ee.es.osaka-u.ac.jp}
\affiliation{%\textit{
${}^1$Center for Quantum Information and Quantum Biology, International Advanced Research Institute, Osaka University, Osaka 560-8531, Japan\\
%Graduate School of Engineering Science, Osaka University, 1-3 Machikaneyama, Toyonaka, Osaka 560-8531, Japan\\
${}^2$Graduate School of Engineering Science, Osaka University, 1-3 Machikaneyama, Toyonaka, Osaka 560-8531, Japan\\
${}^3$RIKEN Center for Quantum Computing (RQC),
Hirosawa 2-1, Wako, Saitama 351-0198, Japan}

 \begin{abstract}
  We introduce SantaQlaus, a resource-efficient optimization algorithm tailored for variational quantum algorithms (VQAs), including applications in the variational quantum eigensolver (VQE) and quantum machine learning (QML). Classical optimization strategies for VQAs are often hindered by the complex landscapes of local minima and saddle points. Although some existing quantum-aware optimizers adaptively adjust the number of measurement shots, their primary focus is on maximizing gain per iteration rather than strategically utilizing quantum shot-noise (QSN) to address these challenges. Inspired by the classical Stochastic AnNealing Thermostats with Adaptive momentum (Santa) algorithm, SantaQlaus explicitly leverages inherent QSN for optimization. The algorithm dynamically adjusts the number of quantum measurement shots in an annealing framework: fewer shots are allocated during the early, high-temperature stages for efficient resource utilization and landscape exploration, while more shots are employed later for enhanced precision. Numerical simulations on VQE and QML demonstrate that SantaQlaus outperforms existing optimizers, particularly in mitigating the risks of converging to poor local optima, all while maintaining shot efficiency. This paves the way for efficient and robust training of quantum variational models.
 \end{abstract}

%\begin{keyword}
%Quantum computing, NISQ, Variational quantum algorithms, Hybrid quantum-classical algorithms, Quantum error mitigation, Effects of the noise
%\end{keyword}

\maketitle
%\onecolumngrid

\section{Introduction}
%Hybrid quantum-classical algorithms is a promising 
%To make use of noisy intermediate-scale quantum (NISQ) devices in the near future \cite{Preskill2018quantumcomputingin}, we have to seek a classically intractable task that hundreds of qubits can resolve under the lack of the error correction.
%A promising framework to realize it is 
%hybrid quantum-classical algorithms, where a classical computer proceeds outputs from a quantum computer which computes some classically intractable functions.
%Especially,
Variational quantum algorithms (VQAs) are promising methods as potential applications for noisy intermediate-scale quantum (NISQ) devices \cite{Preskill2018quantumcomputingin,Cerezo:2021wl}. In VQAs, the loss function is calculated using quantum circuits, and variational parameters of the circuit are optimized via classical computing. Various VQA paradigms have been introduced \cite{Cerezo:2021wl}, including the variational quantum eigensolver (VQE) for ground-state energy approximation \cite{Peruzzo:2014uz,PhysRevX.6.031007,PhysRevX.6.031045,Kandala:2017wb} and others such as QAOA \cite{1411.4028,1602.07674,1712.05771} and quantum machine learning (QML) \cite{PhysRevA.98.032309,Farhi:2018aa,PhysRevA.98.032309,Benedetti_2019,Havlicek:2019ab,Schuld:2019aa,Schuld:2018aa,Schuld:2020aa,Huang:2021aa}.
In QML, quantum neural networks (QNNs) are expected to offer advantages by operating in a feature space that is classically infeasible.
%\cite{Schuld:2018aa,Schuld:2019aa,Havlicek:2019ab,Huang:2021aa}.
While variational (explicit) QNNs can be integrated into quantum kernel methods (implicit model) \cite{Schuld:2021aa}, their distinct importance is underscored by suggestions that they may excel in generalization performance over the kernel methods \cite{Jerbi:2023aa}. When making predictions, quantum kernel models require evaluation of the kernel between new input data and all training data.
 Variational QNNs simply need an evaluation of the input data. Notably, in certain QNN classes, this evaluation can be performed using a classical surrogate \cite{Schreiber:2023aa}. Consequently, the efficient training of such variational models gains significant importance.
%% Barren plateauにも一応触れる。

Classical optimization in VQAs faces multiple challenges that influence the efficiency and reliability of the optimization process.
Barren plateau phenomena, characterized by exponentially vanishing gradients, hinder trainability \cite{McClean:2018aa,Wang:2021wb,Arrasmith:2021aa,Uvarov:2021aa,Sharma:2022aa,Ortiz-Marrero:2021aa,Holmes:2022aa,Holmes:2021aa}.
Likewise, the landscape of local minima and saddle points introduces additional roadblocks for optimization \cite{Bittel:2021aa,Anschuetz:2022aa}.
Although methodologies to mitigate barren plateaus are a topic of ongoing investigation, specific choices of ansatz and cost functions can help in this regard \cite{Cerezo:2021vi,Pesah:2021aa,Zhang:2021aa,Patti:2021aa}.
A core aspect of our research focuses on tackling the issues associated with saddle points and suboptimal local minima as well as efficient resource usage.

Generic classical optimization algorithms, such as Adam \cite{Kingma:2014aa}, Nelder-Mead \cite{Nelder:1965aa}, Powell \cite{Powell:1964aa}, and SPSA \cite{Spall:1987aa} have been widely used in VQAs~\cite{9259985,Kubler2020adaptiveoptimizer,LaRose:2019tb,Havlicek:2019ab,Kandala:2017wb}. In addition to these, quantum-aware optimization schemes have been introduced \cite{Kubler2020adaptiveoptimizer,2004.06252,Stokes2020quantumnatural,PhysRevA.106.062416,nakanishi_sequential_2020,1904.03206,Sweke2020stochasticgradient,PhysRevResearch.4.023017,PRXQuantum.2.030324,2108.10434,2201.13438,Tamiya:2022um,2210.06484,2211.04965,PhysRevX.12.041022,Fontana:2022aa}.
Interestingly, earlier works has noted that low levels of various types of stochastic noise can positively affect the optimization process in VQAs \cite{Sweke2020stochasticgradient,Gentini:2020aa,Patti:2021aa,Patti:2022aa,Duffield:2023aa}.
Especially, in VQAs, the evaluation of expectation values inherently includes statistical noise from quantum measurements, which is called quantum shot-noise (QSN).
Theoretical analysis supports the idea that QSN can assist in escaping the aforementioned optimization traps \cite{Liu:2022aa}.
Nevertheless, current optimization algorithms tend to capitalize on positive effects of QSN only implicitly.

In addition, effective resource allocation during optimization is critical for the practical application of VQAs.
The emphasis should be on effectively utilizing classical data derived from quantum measurements, instead of merely aiming for precise expectation values \cite{Wecker:2015aa,Sweke2020stochasticgradient,Kubler2020adaptiveoptimizer,Tamiya:2022um}.
Existing techniques adjust the number of measurement shots to balance resource use, but often without sufficient regard for the risk of encountering saddle points or converging to poor local optima \cite{Kubler2020adaptiveoptimizer,2004.06252,2108.10434,Tamiya:2022um,Moussa:2023aa,Ito:2023aa}.
 
Motivated by these observations, we address the following question:
Can inherent stochasticity of quantum measurements be strategically leveraged in the optimization process of VQAs in a resource efficient way?
In this paper, to explore this avenue, we propose an optimization algorithm Stochastic AnNealing Thermostats with Adaptive momentum and Quantum-noise Leveraging by Adjusted Use of Shots (SantaQlaus).
SantaQlaus is inspired by a classical optimizer called Stochastic AnNealing Thermostats with Adaptive momentum (Santa) \cite{Chen:2016aa}.
Santa employs simulated Langevin diffusion, guided by an annealing thermostat utilizing injected thermal noise, to approach global optimality.
A key advantage of Santa is its robustness against noise variations, which aligns well with our objectives.

Our main contribution is an extension of the classical Santa optimizer by integrating the leveraging of QSN, resulting in the SantaQlaus algorithm for the optimization of VQAs.
Our proposal seeks to replace the thermal noise in Santa with inherent QSN.
We design SantaQlaus to adaptively adjust the number of shots to ensure the variance of the QSN aligns with the thermal noise utilized in Santa, thus enhancing the efficiency of loss function optimization through annealed thermostats.
Specifically, during the annealing process, fewer shots are required in the early, high-noise stages, while more shots are allocated to the later, low-noise stages, thus demanding more accurate evaluations.
This strategy ensures efficient use of resources without sacrificing accuracy.

Our algorithm is applicable to a wide range of VQAs, especially when the gradient estimator for the loss function shows asymptotic normality. Our method encompasses a general QML framework, accommodating both linear and non-linear dependencies in loss functions, such as mean squared error (MSE), offering flexibility in selecting suitable loss functions for various QML tasks. Additionally, it is compatible with data-independent VQAs like VQE and QAOA.
Indeed, we show the asymptotic normality and its explicit form of mini-batch gradient estimators for linear and quadratic loss functions, extendable to general polynomial loss functions.
From this analysis, we can compute the appropriate number of shots used in our algorithm.

Through numerical simulations on VQE and QML tasks, we demonstrate the superiority of SantaQlaus over established optimizers like Adam and gCANS, showcasing its efficiency in reducing the number of shots required and improving accuracy.
%In this work, ``QSN'' is taken to mean sampling noise in quantum measurements.
%Can we devise a different method to determine the number of shots, one that aggressively leverages QSN to mitigate these risks?
% In this study, we explore this approach by introducing a method that harnesses QSN to circumvent convergence challenges, while also adaptively tuning the shot count. Our proposed optimizer is inspired by the classical Santa optimizer \cite{Chen:2016aa}, which is based on a stochastic-gradient Markov chain Monte Carlo (SG-MCMC) technique.
% We name our optimizer Stochastic AnNealing Thermostats with Adaptive momentum and Quantum-noise Leveraging by Adjusted Use of Shots (SantaQlaus) reflecting the feature of the method.

The remainder of this paper is structured as follows.
Sec.~\ref{s_framework} presents a framework for general QML loss functions in VQAs and the evaluation of the associated gradient estimators.
In Sec.~\ref{s_main}, we provide a comprehensive review of the classical Santa algorithm, which lays the groundwork for the method we introduce.
Sec.~\ref{ss_santa_qlaus} introduces the SantaQlaus algorithm, the principal contribution of this study, beginning with the establishment of the concept of asymptotic normality for the gradient estimators.
Sec.~\ref{sec_numerics} details a series of targeted numerical simulations to assess the performance of the SantaQlaus algorithm, offering empirical evidence to validate its efficacy.
The paper concludes with Section~\ref{jzbmlzhd}, summarizing key findings and suggesting avenues for future research.

\section{Framework}\label{s_framework}
\subsection{Loss functions for VQAs}\label{ss_loss}
Many VQAs are formulated to achieve a goal by minimizing a loss function which is computed by a quantum device with classical parameters to be optimized classically.
In this section, we present a general framework for the loss functions in such VQAs mainly focusing on QML, in a similar manner to Refs.~\cite{Moussa:2023aa,PhysRevA.98.032309}.
We remark that a wide variety of VQAs can be treated in the framework of QML, including VQE and QAOA just by disregarding the data dependency.
The loss functions we explore are formulated to embrace the wide-ranging nature of QML tasks, explicitly accommodating both linear and non-linear dependencies on the expectation values.
As non-linear loss functions are prevalent in machine learning and the choice of loss functions critically influence task performance, it is imperative to ensure the generality of our framework to encompass such functions.

In QML, we have a set of quantum states $\mathcal{S}=\{\rho(\bm{x}_1),\cdots,\rho(\bm{x}_N)\}$ used for training specified by a set of input data $\mathcal{D}=\{\bm{x}_1, \cdots, \bm{x}_N\}$.
Both QC and QQ categories of QML can be put into this formulation, where QC (QQ) aims to learn a classical (quantum) dataset utilizing a quantum algorithm \cite{Aimeur:2006aa,Dunjko:2016aa,Buffoni:2021aa}.
For QC category, the target dataset is the classical data $\mathcal{D}$, and $\mathcal{S}$ is a set of the data-encoded quantum states.
Hence, the choice of the data-encoding feature map $\bm{x} \mapsto \rho(\bm{x})$ from input data $\bm{x}$ into quantum feature $\rho(\bm{x})$ is important, though it is not discussed in detail in this paper.
For QQ purpose, $\bm{x}_i$ is regarded as a specification of each target data quantum state, such as classical descriptions of quantum circuits which generate the states \cite{Nakayama:2023aa}.

In variational QML, a quantum model is given by a parameterized quantum channel $\mathcal{M}_{\bm{\theta}}$, and an observable $H$ to be measured.
In general, we can consider data dependence of the channel as $\rho(\bm{x},\bm{\theta}) = \mathcal{M}_{\bm{\theta}}[\bm{x}](\rho(\bm{x}))$, as in data re-uploading models \cite{Perez-Salinas:2020aa,Schuld:2021ab,Caro:2021aa}.
An observable is given by a hermitian matrix which can be decomposed into directly measured observables $h_j(\bm{x})$ as
\begin{align}
 H(\bm{x}, \bm{w}) = \sum_{j=1}^J w_j c_{j}(\bm{x}) h_j(\bm{x}),
\end{align}
where the observable can be dependent on data $\bm{x}$ and weight vector $\bm{w}$ which is also optimized in general.
Then, the loss function to be minimized for a given task is given as a $p_i$ weighted average
\begin{align}
 L(\bm{\theta}, \bm{w}) = \sum_{i=1}^N p_i \ell(\bm{x}_i, E(\bm{x}_i, \bm{\theta}, \bm{w})),\label{loss_express}
\end{align}
where $\ell$ is a generic function of the data and the expectation value
\begin{align}
 E(\bm{x}, \bm{\theta}, \bm{w}) :=& \Tr \left[\mathcal{M}_{\bm{\theta}}[\bm{x}](\rho(\bm{x})) H(\bm{x}, \bm{w})\right]\nonumber\\
 =& \sum_{j=1}^J w_j c_{j}(\bm{x}) \Tr \left[\mathcal{M}_{\bm{\theta}}[\bm{x}](\rho(\bm{x})) h_j(\bm{x})\right]\nonumber\\
 =:& \sum_{j=1}^J w_j c_{j}(\bm{x}) \langle h_j(\bm{x}) \rangle_{\bm{x}, \bm{\theta}}.
\end{align}
Additionally, some regularization term $\lambda f(\bm{\theta}, \bm{w})$ with a hyperparameter $\lambda$ may be added to the loss function to enhance the generalization performance.
Here, the dependence of the observable on data $\bm{x}$ is considered for the sake of generality.
For example, such a loss function appears in a task to make the output pure state $\rho(\bm{x},\bm{\theta})=\ketbra{\phi(\bm{x},\bm{\theta})}$ of the model close to the correct output state $\rho_{\bm{x}}^{\rm{out}}$, where the loss is given by the average fidelity $\sum_{i=1}^N \bra{\phi(\bm{x}_i,\bm{\theta})}\rho_{\bm{x}_i}^{\rm{out}}\ket{\phi(\bm{x}_i,\bm{\theta})}/N$ \cite{Sharma:2022ab,Beer:2020aa}. This is the case where $\ell(\bm{x},E)=E$, $H(\bm{x})=\rho_{\bm{x}}^{\rm{out}}$ and $p_i=1/N$ without $\bm{w}$.
Variational quantum error correction (VQEC) \cite{Cong:2019ab,Johnson:2017aa} is another example, where the loss function is given by the fidelity between the error corrected (possibly mixed) output state and the ideal pure state.

This framework is applicable to both supervised and unsupervised learning.
In the context of supervised machine learning, we associate each input data point $\bm{x}_i$ with a label $y_i = y(\bm{x}_i)$. It should be noted that the functions $H(\bm{x},\bm{w})$ and $\ell(\bm{x},E)$ implicitly depend on these labels, as they are functions of the data points $\bm{x}$ which include label information via $y(\bm{x})$. This label-dependency will be considered in the definitions and usage of these functions throughout our discussion.
A wide range of the loss functions of VQAs are covered by this form.
A few concrete examples are found in the tasks we employ for numerical simulations in Sec.~\ref{sec_numerics}.
A comprehensive review of various loss functions in QML can be found in Ref.~\cite{Moussa:2023aa}.
For VQAs not involving input data, such as VQE and QAOA, the framework applies by regarding just a single input data to specify the input state being considered \cite{PhysRevA.98.032309}.
In VQE, $\bm{w}$ is fixed and not optimized.

The gradient of the loss function reads
\begin{align}
 \frac{\partial L}{\partial \theta_j} =& \sum_{i=1}^N p_i \frac{\partial \ell}{\partial E} \frac{\partial E}{\partial \theta_j}\label{circ_deriv}\\
 \frac{\partial L}{\partial w_j} =& \sum_{i=1}^N p_i \frac{\partial \ell}{\partial E} \frac{\partial E}{\partial w_j}\label{w_deriv}
\end{align}
by the chain rule.
The derivative with respect to a weight parameter $\frac{\partial E}{\partial w_j}$ is computed as
\begin{align}
 \frac{\partial E}{\partial w_j} = c_j(\bm{x}) \langle h_j(\bm{x}) \rangle_{\bm{x}, \bm{\theta}}.
\end{align}
On the other hand, the derivative $\frac{\partial E}{\partial \theta_j}$ is nontrivial in general.
Because the expectation value is estimated from finite samples of the measurement outcomes, the simple numerical differentiation becomes inaccurate due to the statistical errors \cite{PhysRevLett.126.140502}.
Instead, in our model, we assume that this derivative can be computed by an analytic form via a parameter-shift rule \cite{Wierichs2022generalparameter,PhysRevA.98.032309,PhysRevA.99.032331}
\begin{align}
 \frac{\partial E}{\partial \theta_j} = \sum_{k=1}^{R_j} a_k E(\bm{x}, \bm{\theta} + \epsilon_{j,k}\bm{e}_j, \bm{w}),
\end{align}
where $\bm{e}_j$ is the unit vector in the $j$-th component direction, $a_k$, $\epsilon_{j,k}$ and $R_i$ are constants determined by the model.
In fact, a parameter-shift rule (PSR) holds for a wide range of the model $\mathcal{M}_{\bm{\theta}}$ given by parametric unitary gates.
Especially, if the parameter is given by the gate $U_j(\theta_j) = \exp [-i \theta_j A_j / 2]$ with $A_j^2 = I$, we have \cite{PhysRevA.98.032309,PhysRevA.99.032331}
\begin{align}
\frac{\partial E}{\partial \theta_j} = \frac{E\left(\bm{x}, \bm{\theta} + \frac{\pi}{2} \bm{e}_j, \bm{w}\right) - E\left(\bm{x}, \bm{\theta} - \frac{\pi}{2} \bm{e}_j,\bm{w}\right)}{2}.\label{PSR_2}
\end{align}
In this paper, we assume that two-point PSR (\ref{PSR_2}) holds for all the partial derivatives with respect to $\theta_j$ in our model.

In the following, unless necessary, we omit $\bm{w}$ for brevity.
We call the simplest kind of a loss function with $\ell(\bm{x}, E) = E$ a linear loss function.
Linear loss functions are used in various QML tasks such as quantum auto encoder \cite{Cerezo:2021vi,Romero:2017aa} and VQEC \cite{Cong:2019ab,Johnson:2017aa}, as well as the energy expectation value used in VQE and QAOA.
Non-linear loss functions are also common in QML, such as MSE and the cross entropy (CE) loss functions.
Especially, polynomial loss functions given below are amenable in QML due to the tractability of constructing unbiased estimator:
\begin{align}
 \ell(\bm{x}, E) = \sum_{n=0}^D a_n(\bm{x}) E^n,
\end{align}
where $D$ is the degree of the polynomial.
For MSE given the label $y(\bm{x})$ for the data $\bm{x}$, the function $\ell_{\mathrm{MSE}}$ is given as $\ell_{\mathrm{MSE}}(\bm{x}, E(\bm{x}, \bm{\theta})) = (y(\bm{x}) - \tilde{y}(E(\bm{x}, \bm{\theta})))^2$, where $\tilde{y}(E(\bm{x},\bm{\theta}))$ is a prediction by the model.
Typically, the prediction is given by the expectation value of an observable itself $\tilde{y}(E(\bm{x},\bm{\theta})) = E(\bm{x}, \bm{\theta})$.
In this case, MSE is a kind of the polynomial loss function with $a_0(\bm{x})=y(\bm{x})^2$, $a_1(\bm{x}) = -2y(\bm{x})$, $a_2(\bm{x}) = 1$, and $D=2$.
More general polynomial loss functions are actually used in classical machine learning \cite{Feng:2021aa}.
Ref.~\cite{Feng:2021aa} introduces Taylor-CE, a truncated Taylor series expansion of the CE loss, with the truncation degree serving as a hyperparameter.
Notably, Taylor-CE has been demonstrated to outperform its counterparts in various multiclass classification tasks with label noise, provided that the truncation degree is selected appropriately.

The gradient of a polynomial loss function is given as
\begin{align}
 &\frac{\partial \ell}{\partial E}(\bm{x}, E(\bm{x},\bm{\theta})) \nonumber\\
 =& \sum_{n=1}^D n a_n(\bm{x}) E(\bm{x}, \bm{\theta})^{n - 1}\nonumber\\
 =& \sum_{n=1}^D n a_n(\bm{x}) \left(\sum_j w_j c_j(\bm{x}) \langle h_j(\bm{x}) \rangle_{\bm{x}, \bm{\theta}}\right)^{n - 1}\nonumber\\
 =& \sum_{n=1}^D n a_n(\bm{x}) \nonumber\\
 &\sum_{b_1 + b_2 + \cdots + b_J = n-1}\binom{n-1}{\bm{b}}\prod_j \left(w_j c_j(\bm{x}) \langle h_j(\bm{x}) \rangle_{\bm{x}, \bm{\theta}}\right)^{b_j},
\end{align}
where $\binom{n-1}{\bm{b}}:= \frac{(n-1)!}{b_1!b_2!\cdots b_J!}$.
Thus, computing the derivative with respect to $\theta_j$ ($w_j$) for $D \geq 3$ $(D \geq 2)$ needs an estimate of $\langle h_j(\bm{x}) \rangle_{\bm{x}, \bm{\theta}}^n$ with $n \geq 2$.
For MSE, the derivative with respect to the weight $w_j$ needs an estimate of the squared expectation value.

\subsection{Shot allocation}\label{ss_allocation}
We must decide how to allocate the number of shots to use for the estimation of multiple terms $\langle h_j(\bm{x}) \rangle_{\bm{x}, \bm{\theta}}$, taking into account which terms can be measured simultaneously.
Several strategies have been proposed for the efficient measurement of multiple observables \cite{Huang:2020aa,PhysRevX.6.031007,Kandala:2017wb,Hamamura:2020tc,Crawford2021efficientquantum,PhysRevX.8.031022,Izmaylov:2020wc,Vallury2020quantumcomputed,doi:10.1063/1.5141458,PhysRevA.101.062322,Wu2023overlappedgrouping}.
One simple way is to allocate the shots proportionally to the weight of each simultaneously measurable group deterministically or randomly \cite{2004.06252}.
The weighted random sampling (WRS) achieves an unbiased estimates of the whole loss function even with few number of total shots \cite{2004.06252}, while weighted deterministic sampling (WDS) may be favorable if the observables are grouped into few groups with similar weights.
For simplicity, let each $h_j(\bm{x})$ denote an observable composed of a single group with unit operator norm.
Then, we suppose that the number of shots $s_{i,j}$ to measure $h_{j}(\bm{x}_i)$ is distributed following some probability distribution $P(s_{i,j})$ in general.
Care is needed to estimate a power of an expectation value $\langle h_{j}(\bm{x}_i)\rangle^n$, as the power of a sample average is not an unbiased estimator because powers of each single sample introduce the bias.
In our framework, based on the U-statistic formalism \cite{Moussa:2023aa,Hoeffding:1948aa}, an unbiased estimator of $\langle h_{j}(\bm{x}_i)\rangle^n$ is obtained as
\begin{align}
 \hat{\mathcal{E}}_{i,j,n} = \frac{1}{\mathbb{E}\left[\binom{s_{i,j}}{n}\right]}\sum_{1\leq k_1 < k_2 < \cdots < k_{n} \leq s_{i,j}} \prod_{l=1}^n r_{i,j,k_l},
\end{align}
where $r_{i,j,k}$ denotes the outcome of $k$-th measurement of $h_{j}(\bm{x}_i)$, and $\mathbb{E}[X]$ denotes the expectation value of $X$.

\subsection{Mini-batch gradient}\label{ss_sgd}

This subsection discusses the implementation and benefits of the mini-batch gradient approach in the context of quantum machine learning, contrasting it with the random shot allocation strategy.
A recent optimizer, Refoqus, as mentioned in Ref.~\cite{Moussa:2023aa}, incorporates a unique strategy wherein the number of shots is randomly allocated among data points. This allocation is based on the weight of each term and is an extension of Rosalin \cite{2004.06252} designed for VQE. Consequently, the data points under evaluation are randomly chosen with replacement during the estimation of each gradient component. The number of data points evaluated is autonomously determined through this method.

Despite providing an unbiased gradient estimator that respects weights, this strategy has potential pitfalls for machine learning applications. Firstly, by independently selecting random data points for each gradient component, inter-component gradient correlations are overlooked. As a result, the estimated gradient could be noisier compared to when these correlations are considered. Secondly, choosing data points with replacement means that assessing the entire dataset requires more time than without replacement.
Furthermore, in common scenarios with a uniform weight $p_i=1/N$, uniformly distributing shots across data points in Refoqus often results in the evaluation of a maximal number of data points for the given shot count.
Although such a distribution can minimize the variance of the estimator for a preset shot count, this does not necessarily lead to better optimization performance.
This observation mirrors the fact that stochastic gradient descent often outperforms full-batch gradient descent, even though the latter uses the 'true' gradient.
 In general, the number of data points evaluated at each iteration can have intricate effects on generalization performance, with fewer data points typically yielding superior results.

 Consistent with prevalent machine learning practices, we opt for a mini-batch gradient.
 For simplicity, we consider cases with a uniform weight $p_i = 1/N$.
In a mini-batch strategy, the mini-batch gradient $\nabla \tilde{L}$ of the loss function is evaluated as
\begin{align}
 \frac{\partial \tilde{L}}{\partial \theta_j}(\bm{\theta}) =& \frac{1}{m}\sum_{l=1}^m \frac{\partial \ell}{\partial E}(\bm{x}_{i_l}, E(\bm{x}_{i_l}, \bm{\theta})) \frac{\partial E}{\partial \theta_j}(\bm{x}_{i_l}, \bm{\theta})\nonumber\\
=:& \frac{1}{m}\sum_{l=1}^{m} \mathrm{f}_j(\bm{x}_{i_l},\bm{\theta}).\label{SG_def}
\end{align}
Here, a mini-batch $\{\bm{x}_{i_1}, \cdots, \bm{x}_{i_m}\}$ of size $m$ is chosen randomly from the dataset $\mathcal{D}$ without replacement until the whole dataset is evaluated, at which point it is refreshed.
We allocate an equal number of shots for the evaluation of each data point. In cases where the weight $p_i$ is not uniform, a more effective strategy would be to allocate the number of shots in accordance with this weight.
While this strategy introduces the mini-batch size $m$ as an additional hyperparameter, it can lead to superior performance.
Indeed, in our simulations in Sec.~\ref{sec_numerics}, Refoqus underperforms relative to optimizers that utilize a mini-batch gradient, including the one we propose.

\section{Stochastic annealing thermostats in classical machine learning}\label{s_main}

In this section, as a preliminary to presenting our algorithm, we discuss Santa~\cite{Chen:2016aa}, a classical optimizer that serves as the foundational basis for our approach. Santa was originally developed as an intermediary between stochastic-gradient Markov chain Monte Carlo (SG-MCMC) methods and stochastic optimization, effectively amalgamating the two paradigms.
We start by reviewing the underlying principles of Bayesian sampling and SG-MCMC algorithms, emphasizing their relevance to stochastic optimization, as articulated in Ref.~\cite{Chen:2016aa}.

%As a preliminary to presenting our proposal, we explain a classical optimizer Santa~\cite{Chen:2016aa}, that is the direct parent of our algorithm.
%Santa emerged as a bridge between SG-MCMC and stochastic optimization, embodying a synthesis of these two paradigms.
%We begin by reviewing the Bayesian sampling approach and SG-MCMC algorithms with their relationship to stochastic optimization, drawing from the insights presented in Ref.~\cite{Chen:2016aa}.
%Following this, we detail the classical Santa optimizer \cite{Chen:2016aa}.
%Then, we present our method, incorporating quantum-noise leveraging into Santa.

%\subsection{Bayesian approach with SG-MCMC and stochastic optimizations}
\subsection{Bayesian sampling and stochastic optimizations, SG-MCMC algorithms}\label{ss_classical_santa}

Stochastic optimization methods aim to obtain optimal parameters for an objective function.
Common stochastic optimization methods, such as stochastic gradient descent (SGD), can only find some local minima for non-convex objective functions.
On the other hand, the Bayesian approach provides a probabilistic framework, offering not just point estimates but entire distributions over possible parameter values.
Here, instead of directly finding the optimal parameters, we estimate the likelihood of the parameters given the data.
This probabilistic viewpoint allows for a more exploratory and holistic understanding of the parameter space.

More precisely, the Bayesian approach to machine learning aims to infer the model parameters $\bm{\theta} \in \mathbb{R}^{d}$, from the posterior given by
\begin{align}
 p(\bm{\theta}|\bm{x}_1,\ldots, \bm{x}_N) = \frac{p(\bm{\theta}) \prod_{i=1}^N p(\bm{x}_i|\bm{\theta})}{\int p(\bm{\theta}) \prod_{j=1}^N p(\bm{x}_j|\bm{\theta}) d\bm{\theta}}
\end{align}
upon observing data $\{\bm{x}_1,\ldots, \bm{x}_N\}$. Here, $p(\bm{\theta})$ represents the prior, while $p(\bm{x}_i|\bm{\theta})$ denotes the likelihood of data $\bm{x}_i$ given the model parameter $\bm{\theta}$.
Sampling from the Bayesian posterior distribution offers unique advantages. It allows models to express uncertainty about parameter values, potentially leading to more robust and generalizable solutions.

This task can be equivalently framed as sampling from a probability distribution proportional to $p(\bm{\theta}) \prod_{i=1}^N p(\bm{x}_i|\bm{\theta}) = e^{-L(\bm{\theta})}$, where the negative log-posterior $L(\bm{\theta})$, is defined as 
\begin{align}
 L(\bm{\theta}) = -\log p(\bm{\theta}) - \sum_{i=1}^N \log p(\bm{x}_i|\bm{\theta}).
\end{align}
If $N$ is large, we use a mini-batch stochastic loss function $\tilde{L}_t(\bm{\theta}) := -\log p(\bm{\theta}) - \frac{N}{m}\sum_{j=1}^m \log p(\bm{x}_{j_i}|\bm{\theta})$ in the same way as the stochastic optimization.
The posterior is the same as the Gibbs distribution $\propto e^{-\beta L(\bm{\theta})}$, with the inverse temperature $\beta=1$ and the potential energy $L(\bm{\theta})$.
In fact, the tempered Bayesian posterior $\beta\neq 1$ is recognized as a generalized form of the Bayesian posterior \cite{Vovk:1990aa,McAllester:2003aa,Barron:1991aa,Walker:2001aa,Zhang:2006aa}.
Lowering the value of $\beta$ to less than 1 has been shown to improve the robustness of convergence \cite{Zhang:2003aa}. %Zhang04
This effect has been further explored within the framework of safe-Bayesian methods \cite{Grunwald:2011aa,Grunwald:2012aa,Grunwald:2018aa,Grunwald:2017aa}.
%, which effect has also been studied in the safe-Bayesian framework \cite{}. %Grunwald
Recent research \cite{Wenzel:2020aa,Fortuin:2021aa,Pitas:2022aa} has highlighted the cold posterior effect, wherein sampling from a cold posterior with $\beta > 1$ can offer even better generalization in some scenarios. This approach emphasizes regions of the parameter space that are more consistent with both the prior and the data, providing a nuanced balance between fitting the data and not overfitting.
More generally, the posterior given as the Gibbs distribution $\propto e^{-\beta L(\bm{\theta})}$ for general loss function $L$ is called Gibbs posterior.
%Gibbs sampler Geman
This analogy offers a link between the Bayesian posterior sampling and the stochastic optimization approaches.
Particularly, as $\beta$ tends toward infinity, the procedure converges to the maximum a posteriori (MAP) estimation, aiming for the global minima of the loss function $L$ \cite{Geman:1984aa}.
Thus, sampling from a cold posterior $\propto e^{-\beta L(\bm{\theta})}$ with an elevated $\beta$ guides us closer to minimizing $L$ globally.
%This concept has already appeared in the Gibbs sampler introduced in Geman and Geman's 1984 work \cite{}.

In that sense, the sampling approach can also be applied to optimization of general objective functions $L(\bm{\theta})$ that extend beyond the negative log-posterior, even when they do not have a clear interpretation related to a posterior.
Additionally, it's worth noting that the term $-\log p(\bm{\theta})$ is translated to a regularization term. Indeed, the well-known $L_2$-regularization aligns with the Gaussian prior.

%SG-MCMC algorithms focus on drawing approximate posterior samples of the parameter instead of point estimation of the local minima parameters.
%In classical machine learning, several SG-MCMC algorithms have been proposed \cite{}.
In practice, exact sampling from the posterior is too expensive and we need some approximated sampling method.
SG-MCMC approaches offer efficient sampling methods.
The basic one is the stochastic gradient Langevin dynamics (SGLD) \cite{Welling:2011aa} which updates the parameters as $\bm{\theta}_t = \bm{\theta}_{t-1} - \eta_t \nabla \tilde{L}_t(\bm{\theta}_t) + \sqrt{2\eta_t \beta^{-1}}\bm{\zeta}_t$ with the learning rate $\eta_t$ and the inverse temperature $\beta$, where $\bm{\zeta}_t \sim \mathcal{N}(0, I_d)$ is the additional noise term drawn from the standard normal distribution.
SGLD has been also applied in VQAs \cite{Duffield:2023aa}.
SGLD approximates the (cold) posterior for $\beta=1$ ($\beta > 1$) by simulating the overdamped Langevin dynamics.
The stochastic gradient Hamiltonian Monte Carlo (SGHMC) \cite{Chen:2014aa} incorporates momentum, which corresponds to the underdamped Langevin diffusion.
%SGHMC
Given the recognized importance of momentum in deep model training within stochastic optimization \cite{Sutskever:2013aa}, its incorporation is a logical progression.
 Indeed, SGHMC's connection to SGD with momentum was already highlighted in Ref.~\cite{Chen:2014aa}.
%In the pursuit of higher performance, it is natural to introduce momentum, as it is recognized that using momentum is essential in training deep models for stochastic optimization  \\
%%Stukt
Introducing the friction term in SGHMC is crucial to stabilize the target stationary distribution, preventing stochastic noise from blowing the parameters far away.
Yet, even with this friction term, SGHMC can deviate from the desired thermal equilibrium distribution, particularly when stochastic noise model is poorly estimated.
As remedies, the stochastic gradient Nosé-Hoover thermostat (SGNHT) \cite{Ding:2014aa} and its multivariate counterpart (mSGNHT) \cite{Gan:2015aa} were proposed, adapting the friction term to emulate the energy conservation of the Nosé-Hoover thermostat \cite{Nose:1984aa,Hoover:1985aa}.
%However, SGHMC may still drift away from the target thermal equilibrium distribution if stochastic noise model is poorly estimated.
%Devising the friction term to maintain the kinetic energy similarly to the Nos\'e-Hoover thermostat \cite{}, stochastic gradient Nos\'e-Hoover thermostat (SGNHT) \cite{} and multivariate SGNHT (mSGNHT) \cite{} were introduced. %DingとGan両方

Another important technique is the preconditioning.
In the context of stochastic optimization and SG-MCMC, preconditioning is a technique that aims to improve convergence by appropriately transforming the underlying parameter space, thereby using a metric incorporating a geometric structure that fits the problem.
For a preconditioned SGD \cite{Dauphin:2015aa,Li:2018ac,Li:2018ab}, the update rule is modified by a preconditioner matrix $P_t$ as $\bm{\theta}_t = \bm{\theta}_{t-1} - \eta_t P_t \nabla \tilde{L}_t(\bm{\theta}_t)$.
Adaptive stochastic optimizers such as AdaGrad \cite{Duchi:2011aa}, RMSprop \cite{Tieleman:2012wt}, Adam \cite{Kingma:2014aa}, and their variants
%\cite{Zeiler:2012aa,Carlson:2015aa} 網羅するのはきりがない気がする
incorporate adaptive preconditioning by utilizing historical gradient information to adjust the learning rate for each parameter.
Preconditioning strategies have also found their way into SG-MCMC algorithms \cite{Girolami:2011aa,Patterson:2013aa,Li:2016aa}.
Actually, the significance of judicious preconditioning is widely acknowledged in both realms \cite{Dauphin:2015aa, Girolami:2011aa, Patterson:2013aa,Li:2016aa}.

\subsection{The classical Santa optimizer}\label{ss_cl_santa}
Extending the mSGNHT by incorporating adaptive preconditioning and a simulated annealing scheme on the system temperature, Stochastic AnNealing Thermostats with Adaptive momentum (Santa) was proposed as an optimizer for classical objective functions \cite{Chen:2016aa}.
Santa algorithm is based on a simulated dynamics of the following stochastic differential equation (SDE) given an inverse temperature $\beta$:
\begin{equation}
\left\{
\begin{aligned}
 d\bm{\theta} =& G_1(\bm{\theta}) \bm{p} dt\\
 d\bm{p} =& \left(-G_1(\bm{\theta})\nabla \tilde{L}_t(\bm{\theta}) - \bm{\Xi} \bm{p} \right) dt \\
 &+ \left(\frac{1}{\beta}\nabla G_1(\bm{\theta}) + G_1(\bm{\theta})(\bm{\Xi} - G_2(\bm{\theta}))\nabla G_2(\bm{\theta})\right)dt\\
 &+ \sqrt{\frac{2}{\beta}G_2(\bm{\theta})}d\bm{w}\\
 d\bm{\Xi} =& \left(\mathrm{diag}(\bm{p}^2) - \frac{1}{\beta}I\right) dt
 %d\bm{\Xi} =& \left(\mathrm{diag}(\bm{p}\odot\bm{p}) - \frac{1}{\beta}I\right) dt,
\end{aligned}
       \right.\label{santa_SDE}
\end{equation}
where $\bm{w}$ is the standard Brownian motion, $G_1$ and $G_2$ respectively gives preconditioning for $L(\bm{\theta})$ and the Brownian motion, which encode the respective geometric information.
Here, $\nabla G(\bm{\theta})$ for a matrix $G$ denotes a vector whose $i$-th element is $\sum_{j} \frac{\partial}{\partial \theta_j} G_{i,j}(\bm{\theta})$.
Setting $G_1 = I$ and $G_2$ constant reduces to the SDE for mSGNHT \cite{Gan:2015aa}.
The terms with $\nabla G_1(\bm{\theta})$ and $\nabla G_2(\bm{\theta})$ reflect the spatial variation of the metrics so as to maintain the stationary distribution.
%Using this SDE, we aim to sample the parameters from the target stationary distribution $p_{\beta}(\bm{\theta}) = e^{-\beta L(\bm{\theta})}/Z_{\beta}$ with the normalization factor $Z_\beta$.
Actually, SDE (\ref{santa_SDE}) has the target distribution $p_{\beta}(\bm{\theta}) = e^{-\beta L(\bm{\theta})}/Z_{\beta}$ with the normalization factor $Z_\beta$ as its stationary distribution under reasonable assumptions on stochastic noise as we discuss later.
A remarkable feature of SDE (\ref{santa_SDE}) is that the thermostats $\bm{\Xi}$ maintain the target stationary distribution irrespective of the detail of stochastic noise.
Hence, assuming the ergodicity \cite{Khasminskii:2011aa,Vollmer:2016aa}, we obtain approximate samples from the target distribution $p_{\beta}$ after sufficiently long time.
Then, in the Santa algorithm, the inverse temperature $\beta$ is slowly annealed towards sufficiently large values (hence low temperature) to explore the parameter space to reach near the global optima of the objective function.
After this exploration stage, Santa enters the refinement stage by taking the zero-temperature limit $\beta \rightarrow \infty$ and $\bm{\Xi}$ is fixed to a ``learned value'' in the exploration stage.
The refinement stage is a stochastic optimization with adaptive preconditioning using a learned friction parameter $\bm{\Xi}$.
Indeed, under mild conditions, the asymptotic convergence of Santa algorithm towards the global optima has been shown \cite{Chen:2016aa}.
In fact, it is reported that Santa outperforms conventional optimizers such as SGD, SGD with momentum, SGLD, RMSprop, and Adam in several benchmark tasks \cite{Chen:2016aa}.
Ref.~\cite{Chen:2016aa} tested Santa on training feedforward and convolutional neural networks on the MNIST dataset as well as training recurrent neural network for the task of sequence modeling on four different polyphonic music sequences of piano. In all cited tasks, Santa demonstrates the highest level of performance.
%stationary, ergodicity, annealing, explore, refine (complete explanation of Santa)

To numerically implement the parameter updates according to SDE (\ref{santa_SDE}), we need approximations.
As a simplest Euler scheme introduces relatively high approximation error \cite{Chen:2015aa,Chen:2016aa}, the symmetric splitting scheme (SSS) \cite{Chen:2015aa,Chen:2016aa,Li:2016ab} is recommended.
In Santa, splitting SDE (\ref{santa_SDE}), SSS is implemented by solving each of the following sub-SDEs for finite time steps:
\begin{align*}
&A:\left\{\begin{aligned}
       d\bm{\theta} &= G_1(\bm{\theta}) \bm{p} dt\\
	   d\bm{p} &= 0 \\
	   d\bm{\Xi} &= \left(\mathrm{diag}(\bm{p}^2) - \frac{1}{\beta_t}I\right) dt\\
   \end{aligned},\right.\\
&B:\left\{\begin{aligned}
	   d\bm{\theta} &= 0 \\
	   d\bm{p} &= -\bm{\Xi p}dt\\
       d\bm{\Xi} &= 0\\
   \end{aligned},\right.\\
&O:\left\{\begin{aligned}
       d\bm{\theta} &= 0 \\
	   d\bm{p} &=  -G_1(\bm{\theta})\nabla \tilde{L}_t(\bm{\theta}) dt \\
	   &+ \left(\frac{1}{\beta_t}\nabla G_1(\bm{\theta}) + G_1(\bm{\theta})(\bm{\Xi} - G_2(\bm{\theta}))\nabla G_2(\bm{\theta})\right)dt\\
 &+ \sqrt{\frac{2}{\beta_t}G_2(\bm{\theta})}d\bm{w}\\
       d\bm{\Xi} &= 0.
   \end{aligned}\right.
\end{align*}
For a step size $h$, the solutions are given as
\begin{align}
&A:\left\{\begin{aligned}
       \bm{\theta}_t &= \bm{\theta}_{t-1} + G_1(\bm{\theta}_t) \bm{p}_{t-1} h\\
	   \bm{p}_t &= \bm{p}_{t-1} \\
	   \bm{\Xi}_t &= \bm{\Xi}_{t-1} + \left(\mathrm{diag}(\bm{p}_{t-1}^2) - \frac{1}{\beta_t}I\right) h \\
   \end{aligned},\right.\nonumber\\
&B:\left\{\begin{aligned}
	   \bm{\theta}_t &= \bm{\theta}_{t-1} \\
	   \bm{p}_t &= \exp(-\bm{\Xi}_{t-1}h)\bm{p}_{t-1} \\
       \bm{\Xi}_t &= \bm{\Xi}_{t-1}\\
   \end{aligned},\right.\nonumber\\
&O:\left\{\begin{aligned}
       \bm{\theta}_t &= \bm{\theta}_{t-1} \\
	   \bm{p}_t &= \bm{p}_{t-1} -G_1(\bm{\theta}_t)\nabla \tilde{L}_t(\bm{\theta}_{t}) h \\
	   &+ \left(\frac{1}{\beta_t}\nabla G_1(\bm{\theta}_t) + G_1(\bm{\theta}_t)(\bm{\Xi} - G_2(\bm{\theta}_t))\nabla G_2(\bm{\theta}_t)\right)h \\
 &+ \sqrt{\frac{2}{\beta_t} h G_2(\bm{\theta}_t)}\bm{\zeta}_t \\
       \bm{\Xi}_t &= \bm{\Xi}_{t-1},
   \end{aligned}\right.\label{Santa_update}
\end{align}
where $\bm{\zeta}_t$ is a random vector drawn from $\mathcal{N}(\bm{0}, I)$.
Here, we neglect the change in $G_1(\bm{\theta})$ and $G_2(\bm{\theta})$ within a single parameter update given that $h$ is small enough.
Then, in Santa, parameters are updated in order $A$-$B$-$O$-$B$-$A$ with half steps $h/2$ on $A$ and $B$ updates, and full steps $h$ on the $O$ updates.
%%%%%%%
%%%もともと、nabla G の説明をおいていたばしょ。
%この部分じたいを、SantaQlaus本部に移すほうがよさげ。classical Santa部分ではGの微分に触れない。
% $\bm{\theta}_i = \bm{\theta}_{i-1} + G_1^{i} \bm{p}_{i-1} h$, where $\bm{\theta}_i$, $\bm{p}_i$, and $G_1^{i}$ denote each quantity of $i$-th iteration.

%detail of SDE and noise
SDE (\ref{santa_SDE}) implicitly includes stochastic noise in the mini-batch approximation of the gradient of the loss function $\nabla \tilde{L}_t(\bm{\theta})$.
Based on the central limit theorem, this stochastic noise can be approximated as $G_1(\bm{\theta})\nabla \tilde{L}_t(\bm{\theta}) dt \approx G_1(\bm{\theta})\nabla L(\bm{\theta}) dt + \mathcal{N}(0, 2B(\bm{\theta}) dt)$ with the diffusion matrix $B(\bm{\theta})$ of the stochastic gradient noise \cite{Chen:2014aa}.
Then, for general total diffusion matrix $D(\bm{\theta}) = \frac{1}{\beta}G_2(\bm{\theta}) + B(\bm{\theta})$ including both the injected part $\frac{1}{\beta}G_2(\bm{\theta})$ and stochastic noise part $B(\bm{\theta})$, more precise SDE with the desired stationary distribution becomes \cite{Ding:2014aa}
\begin{align}
 \left\{
\begin{aligned}
 d\bm{\theta} =& G_1(\bm{\theta}) \bm{p} dt\\
 d\bm{p} =& \left(-G_1(\bm{\theta})\nabla L(\bm{\theta}) - \bm{\Xi} \bm{p} \right) dt \\
 &+ \left(\frac{1}{\beta}\nabla G_1(\bm{\theta}) + G_1(\bm{\theta})(\bm{\Xi} - D(\bm{\theta}))\nabla D(\bm{\theta})\right)dt\\
 &+ \sqrt{2 D(\bm{\theta})}d\bm{w}\\
 d\bm{\Xi} =& \left(\bm{p}\bm{p}^{T} - \frac{1}{\beta}I\right) dt.
\end{aligned}
       \right.\label{santa_SDE_precise}
\end{align}
In fact, it has been shown \cite{Ding:2014aa} that the stationary distribution $\pi_{\beta}(\bm{\theta}, \bm{p}, \bm{\Xi})$ of SDE (\ref{santa_SDE_precise}) is given by
\begin{align}
 \pi_{\beta}(\bm{\theta}, \bm{p}, \bm{\Xi}) \propto e^{-\beta L(\bm{\theta}) - \frac{\beta}{2}\bm{p}^T \bm{p} - \frac{\beta}{2}\Tr\left[(\bm{\Xi} - D(\bm{\theta}))^T (\bm{\Xi} - D(\bm{\theta}))\right]}.\label{stationari_whole}
\end{align}
Hence, the marginal stationary distribution of (\ref{stationari_whole}) is the desired one $p_{\beta}(\bm{\theta}) \propto e^{-\beta L(\bm{\theta})}$.
SDE (\ref{santa_SDE}) used in Santa is assured to have the same stationary distribution by assuming the diagonal form of $D(\bm{\theta})$ and neglecting $B(\bm{\theta})$ in the term with $\nabla D(\bm{\theta})$.
This assumption is reasonable when $G_2(\bm{\theta})$ is diagonal and the number of the evaluated data is large enough so that $B(\bm{\theta})$ and its derivatives are small compared to $G_2(\bm{\theta})$ \cite{Ding:2014aa}.
Indeed, only diagonal $G_2(\bm{\theta})$ is used in Santa.
Remarkably, the offdiagonal parts of the SDE for $\bm{\Xi}$ can be omitted in this case, which considerably reduces the computational cost.
%That is reasonable when the injected noise $G_2(\bm{\theta})$
%%%%%% 少し保留。
%Assuming the diagonal form of $D(\bm{\theta})$ $G_2(\bm{\theta})$ and $B(\bm{\theta})$
%We remark that SDE (\ref{santa_SDE}) assumes the diagonal form of $G_2$ and the covariance matrix of stochastic noise in $\nabla \tilde{L}_t(\bm{\theta})$.

The thermostat variable $\bm{\Xi}$ maintains the system temperature by controlling the friction to imposing that the component-wise kinetic energy $p_i^2/2$ is close to $1/(2\beta)$.
If the energy is bigger than this value, the momentum $\bm{p}$ experiences more friction, and the opposite for lower energy values.
Moreover, the thermostat ``absorbs'' the effects of the stochastic term of $D(\bm{\theta})$ \cite{Ding:2014aa}.
In fact, as seen from the stationary distribution (\ref{stationari_whole}), the distribution of $\bm{\Xi}$ is changed to a matrix normal distribution with mean $D(\bm{\theta})$ reflecting stochastic noise $D(\bm{\theta})$, and the marginal stationary distribution of $\bm{\theta}$ is left invariant irrespective of $D(\bm{\theta})$.
As a result, even with imprecise estimation of stochastic noise $B(\bm{\theta})$, we can stably obtain the stationary distribution by the dynamics (\ref{santa_SDE}).
This feature is especially beneficial in its application to VQAs.
%In fact, it has been shown that preconditioning is pivotal for fast and stable convergence in both stochastic optimization \cite{Dauphin:2015aa} and SG-MCMC \cite{Girolami:2011aa,Patterson:2013aa,Li:2016aa}.
%draw advantages from adaptive preconditioners to improve the convergence.

\section{The SantaQlaus algorithm}\label{ss_santa_qlaus}

The classical Santa algorithm obtains (near) global optimality by leveraging annealed thermostats with injected noise.
In VQAs, evaluations of the objective function are inherently noisy due to quantum measurements.
Inspired by this natural intersection of noise characteristics, we propose SantaQlaus, an optimization algorithm designed to leverage intrinsic QSN in VQAs in a resource-efficient manner, emulating thermal noise used in Santa.
Incorporating the asymptotic normality of gradient estimators, our approach justifies the use of QSN as an analogue to thermal noise. Specifically, the asymptotic normality not only provides theoretical underpinning for its use but also guides the adjustment of shot numbers to align with the desired level of thermal noise. 
We start by discussing the asymptotic normality of gradient estimators of loss functions.

\subsection{Asymptotic normality of gradient estimators}\label{ss_normal}
This section delves into the asymptotic normality of gradient estimators for loss functions as a precursor to the deployment of the SantaQlaus algorithm.

Let $\hat{\bm{\mathrm{f}}}(\bm{x}, \bm{\theta})_{\bm{s}} = (\hat{\mathrm{f}}_1(\bm{x}, \bm{\theta})_{\bm{s}},\cdots, \hat{\mathrm{f}}_d(\bm{x}, \bm{\theta})_{\bm{s}})^T$ denote an estimator of the stochastic gradient of loss function $L$ evaluated at a single data point $\bm{x}$ (see Eq.~(\ref{SG_def})) using $\bm{s}=(s_1, \cdots, s_d)$ shots for each component, where $d$ is the number of the parameters including the weight parameters.
We assume a central limit theorem-like asymptotic normality in the following form:
\begin{align}
 \hat{\mathrm{f}}_j(\bm{x}, \bm{\theta})_{\bm{s}}
 =& \mathrm{f}_j(\bm{x}, \bm{\theta}) + \mathcal{N}\left(0, \frac{S_j(\bm{x}, \bm{\theta})}{s_j}\right) + o\left(\frac{1}{\sqrt{s_j}}\right)\nonumber\\
 \approx& \mathrm{f}_j(\bm{x}, \bm{\theta}) + \mathcal{N}\left(0, \frac{S_j(\bm{x}, \bm{\theta})}{s_j}\right)\label{gen_CLT}
\end{align}
with some function $S_j(\bm{x}, \bm{\theta})$, where $\approx$ denotes an approximation up to the leading order terms.
Here, the notation $\mathcal{N}(\mu, \sigma^2)$ is abused to denote a random variable following the normal distribution with mean $\mu$ and variance $\sigma^2$.
A wide range of loss functions actually satisfies this form of asymptotic normality as seen below.
It is worth noting that we do not enforce $s_j$ to be strictly equal to the total number of shots used for estimating each $\mathrm{f}_j(\bm{x}, \bm{\theta})$.
%Here, we do not restrict $s_j$ to exactly equal to the total number of shots used to estimate each $\mathrm{f}_j(\bm{x}, \bm{\theta})$.
Rather, $s_j$ is considered a parameter that characterizes the estimator in such a way that it appears in Eq.~(\ref{gen_CLT}).
%Instead, we rather regard $s_j$ as a parameter characterizing an estimator with Eq.~(\ref{gen_CLT}).
%In our algorithm, we seek an appropriate $s_j$ utilizing Eq.~(\ref{gen_CLT}), and a rule to assign each number of shots for evaluating each expectation values needed to estimate $\mathrm{f}_j$ is prescribed using $s_j$.
A set of rules for determining the number of shots required to evaluate the expectation values needed for estimating $\mathrm{f}_j$ is defined in terms of $s_j$.

For instance, we consider a simple linear loss function with $\ell(E(\bm{x},\bm{\theta}, w)) = w \langle h(\bm{x}) \rangle_{\bm{x},\bm{\theta}}$ given by a single observable $h(\bm{x})$. We have $\mathrm{f}_j(\bm{x},\bm{\theta}, w) = w (\langle h(\bm{x}) \rangle_{\bm{x},\bm{\theta}+ \frac{\pi}{2} \bm{e}_j} - \langle h(\bm{x}) \rangle_{\bm{x},\bm{\theta} - \frac{\pi}{2} \bm{e}_j})/2$ $(j \leq d -1)$ and $\mathrm{f}_d(\bm{x},\bm{\theta}, w) = \langle h(\bm{x}) \rangle_{\bm{x},\bm{\theta}}$, where the weight $w$ is assigned as $d$-th parameter.
Hence, an estimator $\hat{\mathrm{f}}_j(\bm{x}, \bm{\theta})_{\bm{s}}$ is simply given by sample means as
\begin{align}
 \hat{\mathrm{f}}_j(\bm{x}, \bm{\theta}, w)_{\bm{s}} =&
 \frac{w}{2}\left(\sum_{k=1}^{s_j}\frac{r_k^{+}}{s_j} - \sum_{k'=1}^{s_j}\frac{r_{k'}^{-}}{s_j}\right) \quad (j \leq d - 1)\\
 \hat{\mathrm{f}}_d(\bm{x}, \bm{\theta}, w)_{\bm{s}} =&
 \sum_{k=1}^{s_d}\frac{r_k}{s_d},
\end{align}
where $r_k$ and $r_k^{\pm}$ denote the outcome of $k$-th measurement of $h(\bm{x})$ with parameters $\bm{\theta}$ and $\bm{\theta} \pm \frac{\pi}{2}\bm{e}_j$ respectively.
 We note that $2 s_j$ shots are used to estimate $j$-th component because $s_j$ shots are used for each shifted parameter.
 In this case, the central limit theorem reads
\begin{align}
 &\hat{\mathrm{f}}_j(\bm{x}, \bm{\theta}, w)_{\bm{s}}\nonumber\\
 \approx& \mathrm{f}_j(\bm{x}, \bm{\theta}, w) + \mathcal{N}\left(0, w^2 \frac{\sigma^2_{\bm{x},\bm{\theta}+ \frac{\pi}{2} \bm{e}_j} + \sigma^2_{\bm{x},\bm{\theta} - \frac{\pi}{2} \bm{e}_j}}{4 s_j}\right)
 \quad (j \leq d - 1)\nonumber\\
 &\hat{\mathrm{f}}_d(\bm{x}, \bm{\theta}, w)_{\bm{s}}\nonumber\\
 \approx& \mathrm{f}_d(\bm{x}, \bm{\theta}, w)
 + \mathcal{N}\left(0, \frac{\sigma^2_{\bm{x},\bm{\theta}}}{s_d}\right),\label{linear_CLT}
\end{align}
where $\sigma^2_{\bm{x},\bm{\theta}}$ denotes the variance of $h(\bm{x})$ with respect to the state $\rho(\bm{x},\bm{\theta})$.
It is straightforward to generalize Eq.~(\ref{linear_CLT}) to more general cases where we have multiple observables $h_k(\bm{x})$ to evaluate.
We remark that the samples obtained with WRS are equivalent to $s_j$ i.i.d.~samples where each single sample is drawn by measuring a randomly chosen term $h_k(\bm{x})$.
Hence, we can directly apply the above arguments.
%, using randomly distributed numbers of shots, one can apply a central limit theorem for random sums \cite{Marushin:1984aa,Shanthikumar:1984aa}.

For polynomial loss functions, $\mathrm{f}_j$ includes a polynomial of the expectation values of observables in general.
We can apply a central limit theorem for U-statistic \cite{Hoeffding:1948aa} in such cases.
As a typical example, we consider a quadratic loss function $\ell(\bm{x}, E(\bm{x},\bm{\theta}, w)) = \sum_{z=0}^2 a_z(\bm{x}) (w \langle h(\bm{x}) \rangle_{\bm{x},\bm{\theta}})^z$, including the MSE loss for a prediction given by $w \langle h(\bm{x}) \rangle_{\bm{x},\bm{\theta}}$.
In this case, the partial derivatives reads
\begin{align}
 &\mathrm{f}_j(\bm{x},\bm{\theta},w)\nonumber\\
 =& w (\langle h(\bm{x}) \rangle_{\bm{x},\bm{\theta}+ \frac{\pi}{2} \bm{e}_j} - \langle h(\bm{x}) \rangle_{\bm{x},\bm{\theta} - \frac{\pi}{2} \bm{e}_j})\nonumber\\
 &\times\left(\frac{a_1(\bm{x})}{2} + w a_2(\bm{x}) \langle h(\bm{x}) \rangle_{\bm{x},\bm{\theta}}\right) \quad (j \leq d - 1)\\
 &\mathrm{f}_d(\bm{x},\bm{\theta},w)\nonumber\\
 =& a_1(\bm{x})\langle h(\bm{x}) \rangle_{\bm{x},\bm{\theta}} + 2 w a_2(\bm{x}) \langle h(\bm{x}) \rangle_{\bm{x},\bm{\theta}}^2.
\end{align}
Then, for given $\bm{s}$, unbiased estimators of the derivatives are given by
\begin{align}
 \hat{\mathrm{f}}_j(\bm{x}, \bm{\theta}, w)_{\bm{s}} =&
 w\left(\sum_{k=1}^{s_j}\frac{r_k^{+}}{s_j} - \sum_{k'=1}^{s_j}\frac{r_{k'}^{-}}{s_j}\right)\nonumber\\
 &\times
 \left(\frac{a_1(\bm{x})}{2} + w a_2(\bm{x}) \sum_{k=1}^{s_d}\frac{r_k}{s_d}\right)
 \quad (j \leq d - 1)\\
 \hat{\mathrm{f}}_d(\bm{x}, \bm{\theta}, w)_{\bm{s}} =&
 a_1(\bm{x})\sum_{k=1}^{s_d}\frac{r_k}{s_d} + 2 w a_2(\bm{x}) \sum_{1 \leq k_1\neq k_2 \leq s_d}\frac{r_{k_1}r_{k_2}}{s_d(s_d -1)}\label{deriv_d}
\end{align}
with the same notations as above.
In this estimator, we use $s_j$ shots to evaluate at each shifted parameter and $s_d$ shots for the unshifted parameter.
Hence, $2 \sum_{j=1}^{d-1}s_j + s_d$ shots are used in total.
For the circuit parameters $j \leq d - 1$, the estimator composed of a product of sample means of independent random variables.
We note that sequences of two independent and identically distributed (i.i.d.) random variables $X_1,\cdots, X_{n_1}$ and $Y_1, \cdots, Y_{n_2}$ with respective means $\mu_X, \mu_Y$ satisfy
\begin{align}
 &\sum_{k=1}^{n_1}\frac{X_{k}}{n_1}\sum_{k'=1}^{n_2}\frac{Y_{k'}}{n_2} - \mu_X \mu_Y\nonumber\\
 =& \mu_X \left(\sum_{k'=1}^{n_2}\frac{Y_{k'}}{n_2} - \mu_Y\right)
 + \left(\sum_{k=1}^{n_1}\frac{X_{k}}{n_1} - \mu_X\right)\mu_Y\nonumber\\
 &+ \left(\sum_{k=1}^{n_1}\frac{X_{k}}{n_1} - \mu_X\right)\left(\sum_{k'=1}^{n_2}\frac{Y_{k'}}{n_2} - \mu_Y\right).\label{prod_CLT}
\end{align}
Then, we can apply the central limit theorem to $\sum_{k=1}^{n_1}\frac{X_{k}}{n_1} - \mu_X$ and $\sum_{k'=1}^{n_2}\frac{Y_{k'}}{n_2} - \mu_Y$.
 Because the product of the convergent sequences converges to the product of their limits, the third term of the right hand side (RHS) of Eq.~(\ref{prod_CLT}) turns out to be of sub-leading order.
Hence, we obtain
\begin{align}
 \hat{\mathrm{f}}_j(\bm{x}, \bm{\theta}, w)_{\bm{s}}
 \approx
 \mathrm{f}_j(\bm{x}, \bm{\theta}, w)
 + \mathcal{N}\left(0, \frac{S_j(\bm{x},\bm{\theta})}{s_j}\right)
\end{align}
with
 \begin{align}
  &S_j(\bm{x},\bm{\theta})\nonumber\\
=& \mu_{2}^2\left(\sigma^2_{\bm{x},\bm{\theta}+ \frac{\pi}{2} \bm{e}_j} + \sigma^2_{\bm{x},\bm{\theta} - \frac{\pi}{2} \bm{e}_j}\right) + \kappa_j \mu_{1}^2 w^2 a_2(\bm{x})^2 \sigma^2_{\bm{x},\bm{\theta}},\label{prod_var}
 \end{align}
 where $\mu_1 :=  h(\bm{x}) \rangle_{\bm{x},\bm{\theta}+ \frac{\pi}{2} \bm{e}_j} - \langle h(\bm{x}) \rangle_{\bm{x},\bm{\theta} - \frac{\pi}{2} \bm{e}_j}$ and $\mu_2 := \frac{a_1(\bm{x})}{2} + w a_2(\bm{x}) \langle h(\bm{x}) \rangle_{\bm{x},\bm{\theta}}$, and we assume that $s_j/s_d \rightarrow \kappa_j$.
 
 For the weight parameter $j=d$, we apply the theory of U-statistic.
For i.i.d.~random variables $X_1,\cdots, X_n$, U-statistic with respect to a symmetric function $\Phi(\xi_1, \cdots, \xi_m)$ is defined as
 $\sum_{1\leq k_1 \neq \cdots \neq k_m \leq n} \Phi(X_{k_1},\cdots, X_{k_m})/[n(n-1)\cdots (n-m+1)]$.
Thus, Eq.~(\ref{deriv_d}) coincides with the U-statistic with respect to the function $\Phi(\xi_1, \xi_2) = \frac{a_1(\bm{x})}{2}(\xi_1 + \xi_2) + 2 w a_2(\bm{x}) \xi_1 \xi_2$.
Therefore, applying a central limit theorem for U-statistics \cite{Hoeffding:1948aa}, we obtain
\begin{align}
 \hat{\mathrm{f}}_d(\bm{x}, \bm{\theta}, w)_{\bm{s}}
 \approx
 \mathrm{f}_d(\bm{x}, \bm{\theta}, w)_{\bm{s}}
 + \mathcal{N}\left(0, \frac{S_d(\bm{x},\bm{\theta})}{s_d}\right),
\end{align}
where
 \begin{align}
  S_d(\bm{x},\bm{\theta})
  =& \sigma^2_{\bm{x},\bm{\theta}} \left[a_1(\bm{x})^2 + 8 w a_1(\bm{x}) a_2(\bm{x}) \langle h(\bm{x}) \rangle_{\bm{x},\bm{\theta}} \right.\nonumber\\
  &\left. + 16 (w a_2(\bm{x}) \langle h(\bm{x}) \rangle_{\bm{x},\bm{\theta}})^2\right].\label{quad_var2}
 \end{align}
 This way, we obtain the asymptotic normality in a concrete form for quadratic loss functions.
 We can apply similar calculations to more general polynomial loss functions.

 An unbiased estimator $\hat{\bm{\mathrm{f}}}(\bm{\theta})_{\bm{s}}$ of the stochastic gradient $\nabla \tilde{L}(\bm{\theta})$ is given as
 \begin{align}
  \hat{\mathrm{f}}_j(\bm{\theta})_{\bm{s}} := \frac{1}{m}\sum_{l=1}^{m} \hat{\mathrm{f}}_j(\bm{x}_{i_l},\bm{\theta})_{\bm{s}}
 \end{align}
 with respect to a mini-batch $\{\bm{x}_{i_1}, \cdots, \bm{x}_{i_m}\}$ of size $m$.
 Then, we further apply the central limit theorem with respect to the mini-batch average as follows.
 From the asymptotic normality Eq.~(\ref{gen_CLT}), we have
 \begin{align}
  \hat{\mathrm{f}}_j(\bm{\theta})_{\bm{s}} \approx& \frac{1}{m}\sum_{l=1}^{m} \left(\mathrm{f}_j(\bm{x}_{i_l},\bm{\theta}) + \sqrt{\frac{S_j(\bm{x}_{i_l}, \bm{\theta})}{s_j}}\zeta_l\right)\nonumber\\
  =&
  \frac{\partial \tilde{L}}{\partial \theta_j}(\bm{\theta})
  + \frac{1}{m}\sum_{l=1}^m \sqrt{\frac{S_j(\bm{x}_{i_l}, \bm{\theta})}{s_j}}\zeta_{l,j},\label{fj_CLT1}
 \end{align}
 where $\zeta_{l,j}$ $(l=1,\cdots,m), (j=1,\cdots,d)$ are independent standard normally distributed random variables $\sim \mathcal{N}(0, 1)$.
 For simplicity, assuming that $N$ is large enough compared to $m$, we neglect the influence of the sampling without replacement, so that $x_{i_l}$ $(l=1,\cdots,m)$ can be treated as approximately i.i.d.~random variables, which are uniformly drawn from the dataset $\mathcal{D}$.
Indeed, the following arguments based on this approximation can be justified by a central limit theorem for random partition \cite{Li:2017aa,Lehmann:1975aa}.
 Then, $\sqrt{\frac{\bm{S}(\bm{x}_{i_1}, \bm{\theta})}{\bm{s}}}\bm{\zeta}_1, \cdots, \sqrt{\frac{\bm{S}(\bm{x}_{i_m}, \bm{\theta})}{\bm{s}}}\bm{\zeta}_m$ can be regarded as a sequence of i.i.d.~random vectors, whose mean is $\bm{0}$, and the covariance reads
 \begin{align}
  \mathbb{E}\left[\sqrt{\frac{S_j(\bm{X},\bm{\theta})}{s_j}\frac{S_k(\bm{X},\bm{\theta})}{s_k}}\zeta_{j}\zeta_k\right] =& \frac{\mathbb{E}_{\bm{X}}\left[S_j(\bm{X},\bm{\theta})\right]}{s_j}\delta_{j,k}\nonumber\\
  =:& \frac{S_j(\bm{\theta})}{s_j} \delta_{j,k},
 \end{align}
 where $\mathbb{E}_{\bm{X}}$ denotes the expectation with respect to the data sampling $\bm{X}$.
 Here a fraction of vectors mean component-wise.
 Product $\bm{v}_1\bm{v}_2$ of two vectors $\bm{v}_1$ and $\bm{v}_2$ also denote component-wise product in the following.
% We denote $\mathbb{E}_{\bm{X}}\left[S_j(\bm{X},\bm{\theta})\right]$ by $S_j(\bm{\theta})$.
 Then, from Eq.~(\ref{fj_CLT1}), the central limit theorem with respect to the sum $\frac{1}{m}\sum_{l=1}^m \sqrt{\frac{\bm{S}(\bm{x}_{i_l}, \bm{\theta})}{\bm{s}}}\bm{\zeta}_l$ yields an approximation
% \begin{align}
%  \hat{\mathrm{f}}_j(\bm{\theta})_{\bm{s}} \approx&
%  \frac{\partial \tilde{L}}{\partial \theta_j}(\bm{\theta})
%  + \mathcal{N}\left(0, \frac{S_j(\bm{\theta})}{m s_j}\right)
% \end{align}
% for large enough size $m$.
% We also use a vector notation
 \begin{align}
  \hat{\bm{\mathrm{f}}}(\bm{\theta})_{\bm{s}} \approx
 \nabla \tilde{L}_t(\bm{\theta}) + \mathcal{N}\left(\bm{0},\mathrm{diag}\left(\frac{\bm{S}(\bm{\theta})}{m \bm{s}}\right)\right),
 \end{align}
 where $\mathcal{N}(\bm{\mu}, \Sigma)$ denotes a random vector following the multivariate Gaussian distribution with mean $\bm{\mu}$ and covariance matrix $\Sigma$.
 %
% Then, the central limit theorem actually implies that the empirical estimate of the gradient is approximated as
%\begin{align}
% \hat{\bm{\mathrm{f}}}(\bm{\theta})_{\bm{s}} \approx&
% \nabla L(\bm{\theta})\nonumber\\
% &+ \mathcal{N}\left(\bm{0},\mathrm{diag}\left(\frac{\mathbb{E}_{\bm{X}}[\mathbb{V}[\hat{\bm{\mathrm{f}}}(\bm{\theta};\bm{X})|\bm{X}]]}{m \bm{s}}\right)\right)\nonumber\\
% &+ \mathcal{N}\left(\bm{0}, \frac{\Sigma_{\bm{X}}[\mathbb{E}[\hat{\bm{\mathrm{f}}}(\bm{\theta};\bm{X})|\bm{X}]]}{m}\right),\label{emp_CLT}
%\end{align}
%where $\mathbb{E}_{\bm{X}}[\mathbb{V}[\hat{\bm{\mathrm{f}}}(\bm{\theta};\bm{X})|\bm{X}]]$ denotes the vector whose $i$-th component is the expectation value of the conditional variance of $\hat{\mathrm{f}}_i(\bm{\theta};\bm{X})$ given a data $\bm{X}$, with respect to the data $\bm{X}$ as a random variable, and $\Sigma_{\bm{X}}[\mathbb{E}[\hat{\bm{\mathrm{f}}}(\bm{\theta};\bm{X})|\bm{X}]]$ denotes the covariance matrix with respect to $\bm{X}$ of the conditional mean of $\hat{\bm{\mathrm{f}}}(\bm{\theta};\bm{X})$ given a data $\bm{X}$ in a similar way.

\subsection{Details of SantaQlaus}\label{ss_santaQ_algo}
As detailed in the previous section, because samples obtained by quantum measurements are i.i.d.~random variables, we can apply central limit theorems to typical loss functions, so that QSN in a quantum evaluation of each component of the stochastic gradient follows a Gaussian distribution.
 More precisely, with fairly broad applicability, we assume an asymptotic normality Eq.~(\ref{gen_CLT}) of an estimator $\hat{\bm{\mathrm{f}}}(\bm{x}, \bm{\theta})_{\bm{s}}$ of the stochastic gradient of loss function $L$ evaluated at each data point $\bm{x}$.
  Based on this, we have derived an approximation of the stochastic gradient
  \begin{align}
  \hat{\bm{\mathrm{f}}}(\bm{\theta})_{\bm{s}} \approx
 \nabla \tilde{L}_t(\bm{\theta}) + \mathcal{N}\left(\bm{0},\mathrm{diag}\left(\frac{\bm{S}(\bm{\theta})}{m \bm{s}}\right)\right),\label{CLT_main}
  \end{align}
  which is restated for convenience.
  In Eq.~(\ref{CLT_main}), the term $\nabla \tilde{L}_t(\bm{\theta})$ only includes noise due to the mini-batch approximation, and QSN is isolated as Gaussian noise $\mathcal{N}\left(\bm{0},\mathrm{diag}\left(\frac{\bm{S}(\bm{\theta})}{m \bm{s}}\right)\right)$.
Hence, as QSN is approximated as additive Gaussian noise with diagonal covariance matrix, we can use it as thermal noise in Santa.
 We can achieve this by adjusting the number $\bm{n} = m\bm{s}$ to yield the variance corresponding to the desired thermal noise.
% To do so, we have to adjust the variance of QSN which is inversely proportional to the total number of shots  used for each component.
 %%%%% zetaで書いて、方程式の一般型にとどめてごまかす。具体的な場合は、方程式の変形も書く。
% For brevity, the notation $\bm{S}(\bm{\theta}):=\mathbb{E}_{\bm{X}}[\mathbb{V}[\hat{\bm{\mathrm{f}}}(\bm{\theta};\bm{X})|\bm{X}]]$ is used below.
 In the parameter update rule (\ref{Santa_update}) of Santa, with an estimate $\hat{G}_1(\bm{\theta}_t)_{\bm{s}}$ of $G_1(\bm{\theta}_t)$, the stochastic gradient appears as $\hat{G}_1(\bm{\theta}_t)_{\bm{s}} \hat{\bm{\mathrm{f}}}(\bm{\theta}_t)_{\bm{s}} h$, which reads
   \begin{align}
    &\hat{G}_1(\bm{\theta}_t)_{\bm{s}} \hat{\bm{\mathrm{f}}}(\bm{\theta}_{t})_{\bm{s}} h \nonumber\\
   \approx &
  \hat{G}_1(\bm{\theta}_t)_{\bm{s}}\nabla \tilde{L}_t(\bm{\theta}_{t}) h %\nonumber\\
  + h\hat{G}_1(\bm{\theta}_t)_{\bm{s}}\mathrm{diag}\left(\sqrt{\frac{\bm{S}(\bm{\theta}_t)}{\bm{n}}}\right) \bm{\zeta}_t.
   \end{align}
   Comparing with the thermal noise term $\mathcal{N}\left(\bm{0}, \frac{2}{\beta_t} h G_2(\bm{\theta}_t)\right)$, we obtain the following equation to be satisfied for emulating the thermal noise by QSN:
   \begin{align}
    &\hat{G}_1(\bm{\theta}_t)_{\bm{s}}\nabla \tilde{L}_t(\bm{\theta}_{t}) %\nonumber\\
    + \hat{G}_1(\bm{\theta}_t)_{\bm{s}}\mathrm{diag}\left(\sqrt{\frac{\bm{S}(\bm{\theta}_t)}{\bm{n}}}\right) \bm{\zeta}_t\nonumber\\
    =&
    G_1(\bm{\theta}_t)\nabla \tilde{L}_t(\bm{\theta}_{t})
    + \sqrt{\frac{2}{\beta_t h} G_2(\bm{\theta}_t)} \bm{\zeta}_t.\label{variance_eq}
   \end{align}
%   \begin{align}
%    &\mathrm{diag}\left(\frac{\bm{S}(\bm{\theta}_t)}{\bm{n}}\right)\nonumber\\
%    =&
%    \frac{2}{\beta_t h} \hat{G}_1(\bm{\theta}_t)^{-1} G_2(\bm{\theta}_t) (\hat{G}_1(\bm{\theta}_t)^{-1})^T,\label{variance_eq}
%   \end{align}
    %where we assume that $\hat{G}_1(\bm{\theta})$ is invertible.
   We remark that $\hat{G}_1(\bm{\theta}_t)_{\bm{s}}$ may also includes the noise in general.
   Although this condition is not always possible to satisfy for arbitrary $G_1$ and $G_2$, using diagonal preconditioners $G_1$ and $G_2$ makes it possible to solve Eq.~(\ref{variance_eq}) easily.
   Indeed, we use the RMSprop preconditioner and set $G_2(\bm{\theta})h =G_1(\bm{\theta})$ as in the original Santa \cite{Chen:2016aa} for computational feasibility (see Algorithm \ref{alg1}).
   RMSprop preconditioner corresponds to $G_1(\bm{\theta}_t) = \mathrm{diag}(1/\bm{v}_t^{1/4})$, where $\bm{v}_t = \sigma \bm{v}_{t-1} + (1-\sigma)(\nabla \tilde{L}_t(\bm{\theta}_t))^2$ is an exponential moving average of the squared gradient with a parameter $0< \sigma< 1$.
   We leave as future works to incorporate more general preconditioners such as the quantum Fisher metric \cite{Stokes2020quantumnatural,PhysRevA.106.062416}.
   In addition, we assume that $\hat{G}_1(\bm{\theta}_t)$ can be approximated as
   \begin{align}
    \hat{G}_1(\bm{\theta}_t) %\nonumber\\
    \approx G_1(\bm{\theta}_t)\left(I + \mathrm{diag}\left(\bm{g}(\bm{\theta_t})\sqrt{\frac{\bm{S}(\bm{\theta}_t)}{\bm{n}}}\right) \bm{\zeta}_t\right)\label{G1_assump}
   \end{align}
   up to the leading order, where the random vector $\bm{\zeta}_t = \mathcal{N}(\bm{0}, I)$ is shared with the noise in $\hat{\bm{\mathrm{f}}}(\bm{\theta}_t)_{\bm{s}} \approx \nabla \tilde{L}_t(\bm{\theta}_t) + \sqrt{\bm{S}(\bm{\theta}_t)/\bm{n}} \bm{\zeta}_t$.
   In this case, Eq.~(\ref{variance_eq}) reads
    \begin{align}
     & G_1(\bm{\theta}_t)\mathrm{diag}\left(\bm{g}(\bm{\theta}_t)\sqrt{\frac{\bm{S}(\bm{\theta}_t)}{\bm{n}}}\right)\nabla \tilde{L}_t(\bm{\theta}_{t}) \bm{\zeta}_t\nonumber\\
    &+G_1(\bm{\theta}_t)\mathrm{diag}\left(\sqrt{\frac{\bm{S}(\bm{\theta}_t)}{\bm{n}}}\right) \bm{\zeta}_t\nonumber\\
    =&
    \sqrt{\frac{2}{\beta_t h} G_2(\bm{\theta}_t)} \bm{\zeta}_t.
    \end{align}
   Thus, substituting $G_2(\bm{\theta}) =G_1(\bm{\theta})/h$, we obtain the appropriate number of shots as follows:
   \begin{align}
    &\bm{n} \nonumber\\
    =& \left\lceil\frac{\beta_t h^2}{2} G_1(\bm{\theta}_t) \left(1 + \bm{g}(\bm{\theta}_t)\nabla \tilde{L}_t(\bm{\theta}_{t})\right)^2 \bm{S}(\bm{\theta}_t)\right\rceil.\label{SQ_shots_rule}
   \end{align}
   In particular, by neglecting the noise in $\bm{v}_{t-1}$ of the previous iteration, $G_1$ of the RMSprop preconditioner actually satisfies Eq.~(\ref{G1_assump}) since it is given by an estimate of the squared gradient.
   For RMSprop preconditioner, we have
   \begin{align}
    &\hat{G}_1(\bm{\theta}_t) \nonumber\\
    \approx&
    \mathrm{diag}\left(\left(\sigma \bm{v}_{t-1} + (1-\sigma)\hat{\bm{\mathrm{f}}}(\bm{\theta}_t)_{\bm{s}}^2\right)^{-\frac{1}{4}}\right)\nonumber\\
    \approx&
    {G}_1(\bm{\theta}_t)
    \left(I + 2(1-\sigma)\mathrm{diag}\left(\frac{\nabla \tilde{L}_t(\bm{\theta}_{t})}{\bm{v}_t}\sqrt{\frac{\bm{S}(\bm{\theta}_t)}{\bm{n}}}\right)\bm{\zeta}_t\right)^{-\frac{1}{4}}\nonumber\\
    \approx&
    {G}_1(\bm{\theta}_t)
    \left(I - \frac{(1-\sigma)}{2}\mathrm{diag}\left(\frac{\nabla \tilde{L}_t(\bm{\theta}_{t})}{\bm{v}_t}\sqrt{\frac{\bm{S}(\bm{\theta}_t)}{\bm{n}}}\right)\bm{\zeta}_t\right),
   \end{align}
   where we use the Taylor expansion and neglect the terms of sub-leading order.
   Hence, $\bm{g}(\bm{\theta}_t) = - 0.5 (1-\sigma)\nabla \tilde{L}_t(\bm{\theta}_{t})/\bm{v}_t$ holds.
   Then, estimating the gradient using the number of shots given by Eq.~(\ref{SQ_shots_rule}), we apply parameter update rule (\ref{Santa_update}) of Santa.
   %without injected noise. %%%%%%%%% noise injection
   We call this optimization algorithm SantaQlaus, which stands for Stochastic AnNealing Thermostats with Adaptive momentum and Quantum-noise Leveraging by Adjusted Use of Shots.
   A small number of shots is sufficient when the temperature is high, and as the optimization proceeds and the temperature is lowered, the number of shots required is increased.
   This allows us to efficiently leverage QSN to explore the parameter space in accordance with the Langevin diffusion while saving the number of shots.
   %%%%%%%%%%
%   In this way, we can efficiently leverage QSN as the thermal noise.
   %In accordance with the annealing schedule $\beta_t$, we can emulate the thermal noise by adjusting the number of shots
   Though the exact evaluation of $\bm{S}(\bm{\theta}_t)$ is infeasible, we can estimate it by computing the mini-batch average of an estimator of $\bm{S}(\bm{x_{i_l}},\bm{\theta})$.
   For linear loss function, we can estimate $\bm{S}(\bm{x_{i_l}},\bm{\theta})$ by the unbiased variance.   
   For general polynomial loss functions, we can estimate it via corresponding U-statistics.
   In particular, it is straightforward to calculate the U-statistics to estimate Eq.~(\ref{prod_var}) and (\ref{quad_var2}) for quadratic loss functions including MSE.
   However, we should note that an unbiased estimator given by a U-statistic is not guaranteed to be positive for general cases.
   If an estimate of $\bm{S}(\bm{x_{i_l}},\bm{\theta})$ is negative, we cannot use it to calculate the number of shots.
   In such a case, when the obtained estimator takes a negative value, we switch to use a biased estimator guaranteed to be positive, which is obtained by simply substituting corresponding sample means (unbiased variance) to expectation values (variance).
  In addition, we use the exponential moving average as estimates of the quantities for the next iteration similarly to \cite{Kubler2020adaptiveoptimizer,2108.10434}.
   Of course this approach is subject to errors resulting from taking the ceil and estimation errors, but the thermostat absorbs their effects, as seen in the previous section.
   We remark that the overhead of estimating $\bm{S}(\bm{\theta}_t)$ is only classical calculation of the statistics using the same samples of quantum outputs as those used in estimating the gradient.
%Here and thereafter, fractional operations between vectors mean component-wise.
%In fact, QSN can be treated similarly to the injected noise in contrast to stochastic noise resulting from the mini-batch approximation.
%stochastic noise in the mini-batch gradient is poorly controllable 
%Although stochastic noise resulting from the mini-batch approximation is also present in the stochastic gradient of classical loss functions, it is poorly controllable.
%First, the estimation of the variance of the noise in the mini-batch approximation is difficult, as
%Second, the freedom in choosing a mini-batch size is limited not only because using a large mini-batch size is costly but also because the choice of mini-batch size affects generalization performance.
%このあたり、ミニバッチのノイズについての説明は正確に言うのが難しいので保留
%%古典でなぜそうできなかったか、と、量子ではうまくあっていることの説明。成分独立とショット数の自由さ。データ点の数は質的に違う
%In fact
Santa algorithm is especially a good fit for leveraging QSN this way.
This strategy can be straightforwardly applied to other VQAs without input data such as VQE, not only limited to QML.

We note that the variance of QSN cannot exceed that at the minimum number of shots.
An option to address this constraint, as well as the variance estimation error, is to inject noise with the missing variance.
Although this approach might potentially enhance the ergodicity of the dynamics, our numerical experiments did not validate the benefit of this option, as it primarily delayed the optimization.
The pitfalls seem to eclipse the advantages, given there is no necessity for precise sampling at the outset and the appropriate variance of the noise to inject is challenging to estimate accurately.
%We believe that the disadvantages outweighed the advantages, since there is no need to sample precisely initially and the variance of the noise to be injected cannot be estimated precisely.

The convergence theorem towards the global optima under mild conditions for Santa \cite{Chen:2016aa} remains valid for SantaQlaus.
Even if the actual variance of QSN does not exactly yield thermal noise $\frac{2}{\beta_t} G_2(\bm{\theta}_t)\bm{\zeta}_t$ to be emulated, we can redefine $G_2(\bm{\theta})$ so that thermal noise with it coincides with actual QSN.
In this way, we obtain the same SDE as the original Santa with this redefined $G_2(\bm{\theta})$.
Thus, we can apply convergence theorem \cite[Theorem 2]{Chen:2016aa} to SantaQlaus to guarantee its convergence.
Moreover, even though the approximation by the central limit theorem might be imprecise due to the limited shots in early iterations, regarding an iteration with a sufficient number of shots as the starting point ensures that the asymptotic convergence behavior remains unaffected.
 Practically speaking, the sampling accuracy in initial iterations does not significantly influence the optimization.
 % While there may be some benefit in increasing the ergodicity of the dynamics by doing this, we could not confirm the advantage of this option in our numerical experiments, as it only delay the optimization.

Another remark is drawn for the application of SantaQlaus to quadratic loss functions.
As it is impractical to set the ratio $\kappa_j = s_j/s_d$ in Eq.~(\ref{prod_var}) a priori, we instead enforce $\kappa_j \leq 1$ and use an upper bound $\bar{S}_j(\bm{x},\bm{\theta}) := \mu_{2}^2\left(\sigma^2_{\bm{x},\bm{\theta}+ \frac{\pi}{2} \bm{e}_j} + \sigma^2_{\bm{x},\bm{\theta} - \frac{\pi}{2} \bm{e}_j}\right) + \mu_{1}^2 w^2 a_2(\bm{x})^2 \sigma^2_{\bm{x},\bm{\theta}} \geq S_j(\bm{x},\bm{\theta})$ to determine $s_j$.
We can do so by first computing $s_j$ using $\bar{S}_j(\bm{x},\bm{\theta})$ $(j \leq d - 1)$, and $s_d$ via $S_d(\bm{x},\bm{\theta})$.
Then, we use $\max\{s_1,\cdots, s_d\}$ shots to measure $h(\bm{x})$ at the unshifted parameter $\bm{\theta}$.
In this way, we can adjust the variance of QSN to be at least smaller than that of the desired thermal noise.

%%%%%%%%to modify
%Although the approximation by the central limit theorem may not be precise because of the small number of shots in early iterations, regarding the iteration at which a sufficient number of shots is used as the initial step, we can see that the asymptotic convergence behavior is not affected.
%In practice, too, the accuracy of sampling in the early iterations does not have much impact on the optimization.

\begin{figure}[!t]
\begin{algorithm}[H]
    \caption{SantaQlaus (a practical implementation). The function $iEvaluate(\bm{\theta}, \bm{s}, m)$ evaluates the mini-batch gradient estimator $\bm{\mathrm{f}} = \hat{\bm{\mathrm{f}}}(\bm{\theta})_{\bm{s}}$ for the objective loss function with a size $m$ mini-batch, using the number of shots determined by $\bm{s}$. This function also returns an estimator $\bm{S}$ of $\bm{S}(\bm{\theta})$ obtained by taking mini-batch average of the corresponding U-statistic computed from the measurement outcomes. The regularization term (or a prior) can also be specified in $iEvaluate$. The function $sCount(\bm{s}, m)$ returns the total number of shots expended in $iEvaluate(\bm{\theta}, \bm{s}, m)$. For example, for a linear loss function whose gradient is computed by two-point parameter shift rule (\ref{PSR_2}), $sCount(\bm{s}, m) = 2 m \sum_{i} s_i$. We can also straightforwardly apply this algorithm to a VQA without data dependence such as VQE just by neglecting the argument $m$ and setting $m=1$.}
    \label{alg1}
    \begin{algorithmic}[1]
    \REQUIRE Learning rate $\eta_t$, initial parameter $\bm{\theta}_0$, minimum number of shots $s_{\min}$, total shot budget $s_{\max}$ available for the optimization, annealing schedule $\beta_t$, mini-batch size $m_t$, scale factor of the preconditioner $\bm{g}_t$, warm-up iteration number $t_0$, hyperparameters $\sigma$, $C$, and $\lambda$, running average constant $\mu$ %, burn-in rate $r$
    %\ENSURE $y = x^n$
    \STATE Initialize: $\bm{\theta} \leftarrow \bm{\theta}_0$, $t \leftarrow 1$, $s_{\mathrm{tot}} \leftarrow 0$, $\bm{s} \leftarrow (s_{\min},\cdots,s_{\min})^{\mathrm{T}}$, $\bm{\xi}' \leftarrow (0,\cdots,0)^{\mathrm{T}}$, $\bm{\chi}' \leftarrow (0,\cdots,0)^{\mathrm{T}}$, $\bm{\Gamma}' \leftarrow (0,\cdots,0)^{\mathrm{T}}$, $\bm{v} \leftarrow (0,\cdots,0)^{T}$, $\bm{u} \leftarrow \sqrt{\eta_1} \mathcal{N}(0, I)$, $\bm{\alpha} \leftarrow \sqrt{\eta_1} C$
    %\IF{$n < 0$}
    %\STATE $X \leftarrow 1 / x$
    %\STATE $N \leftarrow -n$
    %\ELSE
    %\STATE $X \leftarrow x$
    %\STATE $N \leftarrow n$
    %\ENDIF
    \WHILE{$s_{\mathrm{tot}} \leq s_{\max}$}
     \STATE $\bm{\mathrm{f}}, \bm{S} \leftarrow iEvaluate(\bm{\theta}, \bm{s}, m_t)$
     \STATE $s_{\mathrm{tot}} \leftarrow s_{\mathrm{tot}} + sCount(\bm{s}, m_t)$
     \STATE $\bm{v} \leftarrow \sigma \bm{v} + (1 - \sigma) \bm{\mathrm{f}}^2$
     \STATE $\bm{G} \leftarrow \bm{g}_t/\sqrt{\lambda + \sqrt{\bm{v}}}$
     \STATE $\bm{\theta} \leftarrow \bm{\theta} + \bm{G} \bm{u}/2$
     \STATE $\bm{\alpha} \leftarrow \bm{\alpha} + (\bm{u}^2 - \eta_t/\beta_t)/2$
     \STATE $\bm{u} \leftarrow \exp(-\bm{\alpha}/2)\bm{u}$
     \STATE $\bm{u} \leftarrow \bm{u} - \eta_t \bm{G}\bm{\mathrm{f}}$
     \STATE $\bm{u} \leftarrow \exp(-\bm{\alpha}/2)\bm{u}$
     \STATE $\bm{\alpha} \leftarrow \bm{\alpha} + (\bm{u}^2 - \eta_t/\beta_t)/2$
     \STATE $\bm{\theta} \leftarrow \bm{\theta} + \bm{G} \bm{u}/2$
     \STATE $t \leftarrow t + 1$
     \IF{$t > t_0$}
     \STATE $\bm{\xi}' \leftarrow \mu \bm{\xi}' + (1 - \mu)\bm{S}$
     \STATE $\bm{\xi} \leftarrow \bm{\xi}' / (1 - \mu^{t-t_0})$
     \STATE $\bm{\chi}' \leftarrow \mu \bm{\chi}' + (1 - \mu)\bm{\mathrm{f}}$
     \STATE $\bm{\chi} \leftarrow \bm{\chi}'/ (1 - \mu^{t - t_0})$
     \STATE $\bm{\Gamma}' \leftarrow \mu\bm{\Gamma}' + (1 - \mu) \bm{G}$
     \STATE $\bm{\Gamma} \leftarrow \bm{\Gamma}' / (1 - \mu^{t-t_0})$
     \STATE $\bm{v}' \leftarrow \sigma \bm{v} + (1-\sigma) \bm{\chi}$
     \STATE $\bm{\gamma} \leftarrow \left(1 - 0.5 (1-\sigma)\bm{\chi}^2/\bm{v}'\right)^2$
     \STATE $\bm{n} \leftarrow \left\lceil \beta_t \eta_t \bm{\Gamma}\bm{\gamma} \bm{\xi}/2 \right\rceil$
     \STATE $\bm{s} \leftarrow \left\lceil \bm{n} / m_t \right\rceil$
     \STATE $\bm{s} \leftarrow \mathrm{clip}(\bm{s}, s_{\min}, \mathrm{None})$
     \ELSE
     \STATE $\bm{s} \leftarrow (s_{\min},\cdots,s_{\min})^{\mathrm{T}}$
     \ENDIF    
    \ENDWHILE
    \end{algorithmic}
\end{algorithm}
\end{figure}

A practical implementation of our SantaQlaus algorithm is summarized in Algorithm \ref{alg1}.
Here, we re-parameterize as $\eta = h^2$, $\bm{u} = \sqrt{\eta}\bm{p}$, and $\mathrm{diag}(\bm{\alpha}) = \sqrt{\eta}\bm{\Xi}$ as in Ref.~\cite{Chen:2016aa,Ding:2014aa}.
We do not use the update rule of the refinement stage because we cannot make QSN exactly zero in contrast to the classical case.
Instead, we can incorporate the refinement stage into the annealing schedule of $\beta_t$ in such a way that a large enough value of $\beta_t$ is used in the refinement.
We detail an example of such a strategy later.
One obstacle in an implementation of Santa's update rule is the terms with $\nabla G_1(\bm{\theta})$ and $\nabla G_2(\bm{\theta})$ in SDE (\ref{santa_SDE}).
Exact calculation of $\nabla G_1(\bm{\theta})$ and $\nabla G_2(\bm{\theta})$ is infeasible in general.
As shown in Ref.~\cite{Chen:2016aa}, one approach is to approximate them by applying Taylor expansion with respect to the parameter update.
This approach actually yields a computationally efficient approximation.
However, in practice, this approximation can be numerically unstable due to the time discretization and rapid changes in some components of the preconditioner.
As a remedy, it was empirically found beneficial to simply drop out the terms with $\nabla G_1$ and $\nabla G_2$ \cite{Chen:2016aa}.
Even if this is done, slightly biased samples do not affect the optimization so much as our purpose is not to obtain the accurate sampling from the target distribution.
In fact, such a small bias is absorbed into the error in the estimated stochastic noise.
Even for a sampling task, it has been shown that the bias caused by neglecting the derivatives of the preconditioner can be controlled to be small for the RMSprop preconditioner \cite{Li:2016aa}.
Then, we also neglect the terms with $\nabla G_1$ and $\nabla G_2$ in a practical implementation, just as done in the classical Santa \cite{Chen:2016aa}.

Additionally, we introduce a warm-up iteration number, denoted as $t_0$, and a component-wise scale factor, $\bm{g}_t$, for the preconditioner to enhance flexibility.
During the very early iterations, the estimation for quantities such as $\bm{S}$ and $\bm{G}$ can be unstable, with some components disproportionately large.
Incorporating such unstable values into the moving average can be detrimental.
Consequently, the number of shots $\bm{n}$ computed from these early estimates may be erratic.
To counteract this, it is advantageous to disregard these values in a few early iterations.
We might choose a small value for $t_0$, such as $5$.
While the default for $\bm{g}_t$ can be set to 1, when components have varying optimal scales, adjusting the learning rate individually for each component is beneficial.
For instance, in regression tasks of QML with a scaled observable, as discussed in Sec.~\ref{ss_regression}, allowing the scale parameter to adjust quickly is favorable as it significantly influences the success of the regression.
However, introducing a component-wise different learning rate $\bm{\eta}_t$ is not justified because $\eta_t$ represents time.
We empirically found that such modifications degrade the performance.

%%%%%%%%%%%%%%%%

As the shot budget is important, schedules of hyperparameter settings of such as $\beta_t, \eta_t$, or $m_t$ based on the number of shots may be useful.
For instance, we use the following function of the expended shots $s_{\mathrm{tot}}$ to determine the value of the hyperparameters to be used:
\begin{align}
 f_{s_0,s_{\mathrm{end}}}^{y_0,y_{\mathrm{end}},a}(s_{\mathrm{tot}}) := y_{0} \left[\frac{s_{\mathrm{tot}}-s_{0}}{s_{\mathrm{end}}-s_{0}}\left(
 \left(\frac{y_{\mathrm{end}}}{y_0}\right)^{\frac{1}{a}} - 1\right) + 1\right]^a.\label{stot_beta}
\end{align}
This function takes the predetermined values $y_0$ and $y_{\mathrm{end}}$ at the start $s_0$ and the end $s_{\mathrm{end}}$ respectively, and the curve of the growth is controlled by $a\neq 0$.

\subsection{Incorporating quantum error mitigation}\label{hardware_noise}

In the context of VQAs executed on NISQ devices, addressing hardware noise is important.
Quantum error mitigation (QEM) techniques offer a pathway for estimating the output of an ideal, noiseless circuit from noisy ones \cite{PhysRevX.7.021050,Endo:2021aa,PhysRevLett.119.180509,PhysRevX.8.031027,Czarnik:2021aa,Strikis:2021aa,Berg:2023aa,Kandala:2019aa,Cai:2021aa,Mari:2021aa,Nation:2021aa,Yang:2022aa,Cai:2022aa}, without large resource overhead required in quantum error correction.
In principle, it is possible to consider the noisy ansatz instead as our model.
However, to obtain an accurate result, we finally need to remove the effects of the hardware noise at least for the evaluation of the final outputs after the optimization, highlighting the need for QEM in real devices \cite{Kandala:2017wb,Kandala:2019aa,Kim:2023aa}.
Thus, we should aim to optimize the parameters with respect to the noiseless model.
While some VQA tasks exhibit resilience to hardware noise \cite{Gentini:2020aa,Sharma:2020aa,PhysRevA.104.022403},
the presence of such noise can introduce computational bias in gradient estimation, thereby altering the loss landscape and complicating optimization in general \cite{PhysRevA.104.022403,Wang:2021wb}.
 Prior research indicates that employing QEM during optimization can potentially enhance the performance \cite{Havlicek:2019ab,Wang:2021aa,Jose:2022aa,Wang:2022aa}, although QEM is unlikely to resolve the noise-induced barren plateau \cite{Wang:2021wb,Wang:2021aa}.
However, it is not always the case, as the application of QEM inherently introduces a trade-off: while it reduces bias, it increases the sampling variance \cite{Wang:2021aa,Quek:2022aa,Tsubouchi:2022aa,Takagi:2022ab,Takagi:2022aa}.
 For instance, probabilistic error cancellation (PEC) effectively applying the inverse map of the noise channel via sampling the circuits according to a quasi-probability decomposition of the inverse map \cite{PhysRevLett.119.180509,PhysRevX.8.031027,Wang:2022aa}.
As a result, it yields an unbiased estimator of the noiseless expectation value at the expense of increased variance.
 Zero noise extrapolation (ZNE) \cite{PhysRevX.7.021050,PhysRevLett.119.180509,PhysRevX.8.031027} also yields a bias-reduced estimator with the increased variance.
 
 The integration of QEM techniques into SantaQlaus may offer a resource-efficient use of these methods during optimization.
 For example, PEC can be seamlessly incorporated into the SantaQlaus framework.
PEC provides an unbiased estimator for the gradient, enabling optimization of the noiseless model effectively \cite{Jose:2022aa}.
The distinction lies in the increased number of circuit terms that must be sampled, as dictated by the quasi-probability decomposition of the inverse noise map.
In SantaQlaus, an appropriate total number of measurement shots can be computed from the gradient estimator's variance $\bm{S}(\bm\theta)$, obtained from the PEC samples.
These shots are then allocated based on the quasi-probability distribution.
This approach enables SantaQlaus to operate using an unbiased gradient in the presence of hardware noise via PEC, while also automatically adjusting resource usage for efficiency.
In a sense, SantaQlaus may also leverage hardware noise that is ``converted'' to sampling noise with increased variance via QEM.

Similarly, ZNE can be incorporated into SantaQlaus to provide a bias-reduced estimator of the gradient.
Although the estimator is not perfectly unbiased in ZNE, such a bias-reduced estimator may still offer satisfactory performance in optimizing variational parameters, as observed in an experiment \cite{Havlicek:2019ab}.
Other QEM methods such as Clifford data regression \cite{Czarnik:2021aa} can be incorporated in a similar manner.
This way, SantaQlaus has a potential to utilize QEM in a resource-efficient manner.

\section{Numerical simulations}\label{sec_numerics}
In this section, we demonstrate the performance of SantaQlaus optimizers against a range of existing optimizers through numerical simulations of a benchmark VQAs.
For the optimization process, we simulate the sampling of the outcomes of quantum measurements of loss functions via  Qulacs \cite{Suzuki2021qulacsfast}.
The resulting optimization curves are plotted based on the exact evaluations via state vector calculations provided by Qulacs.
We do not include hardware noise.

We benchmark SantaQlaus against several other optimization algorithms.
As a representative of generic optimizers, we employ Adam \cite{Kingma:2014aa}, renowned for its broadly effective performance.
To test the advantages of adaptive shots-adjusting strategy of SantaQlaus over simpler methods, we use Adam with a predetermined increased number of shots, denoted as Adam with dynamic shots (Adam-DS).
In addition, we compare SantaQlaus with existing shot-adaptive optimizers, such as gCANS \cite{2108.10434} and its QML-specific variant, Refoqus \cite{Moussa:2023aa}.
%As for an existing shot-adaptive optimizer, we compare gCANS \cite{2108.10434} and its QML specific version Refoqus \cite{Moussa:2023aa}.

 As another optimizer that aims for global optimality with MCMC approach, we pick up MCMC-VQA \cite{Patti:2022aa}.
 It is based on Metropolis–Hastings steps using injected stochastic noise, which can be resource-intensive to achieve sufficient mixing.
 The shot resource efficiency of MCMC-VQA has not been comprehensively studied.
 We also compare SantaQlaus with MCMC-VQA in a VQE task (Sec.~\ref{ss_TFIM}).

%In QML benchmarks, we compare SantaQlaus with Adam-DS and 
% Refoqus \cite{Moussa:2023aa} and Adam \cite{Kingma:2014aa} with gradually increased mini-batch size and the number of shots according to a similar schedule to (\ref{stot_beta}).

 %The minimum number of shots  is used for we-AdamCANS and AdamCANS.
 We fix a part of the hyperparameters to recommended values throughout the benchmarks.
 For SantaQlaus, we use $\sigma=0.99$, $C=5$, $\lambda=10^{-8}$, $\mu=0.99$, $\eta_1 = 0.01$, $s_{\min}=4$, and $t_0 = 5$.
  For Adam, $\beta_1=0.9$, $\beta_2 = 0.99$, $\epsilon = 10^{-8}$ are used.
  In MCMC-VQA, we use the inverse temperature $\beta = 0.2$ and the noise parameter $\xi = 0.5$ as recommended in Ref.~\cite{Patti:2022aa}.
%For the other hyperparameters, the default values of each optimizer given in each proposal paper are used.
We do not use artificial noise injection in SantaQlaus since we could not find any improvements.
For each optimizer, we choose the best hyperparameters in a grid search.
For all scheduled hyperparameters, we use the function $f_{s_0,s_{\mathrm{end}}}^{y_0,y_{\mathrm{end}},a}(s_{\mathrm{tot}})$ given in Eq.~(\ref{stot_beta}) to assign the value according to the expended shots.
 For the performance with respect to the shots resource usage, we empirically found this strategy is better than the usual one based on a function of the number of the iteration.

 We use a gradually decreasing learning rate $\eta_t = f_{0,s_{\max}}^{\eta_1,\eta_{\mathrm{end}},a_{\mathrm{LR}}}(s_{t})$, where $s_t$ denotes $s_{\mathrm{tot}}$ up to $t$-th iteration.
 For gCANS and Refoqus, the fixed learning rate $1/L$ is used with the Lipschitz constant $L$ of the gradient.
 Throughout the benchmarks and the other optimizers, we use $\eta_1=0.01$ and $\eta_{\mathrm{end}} = 0.001$.
 The exponent $a_{\mathrm{LR}}$ is chosen for each benchmark and each optimizer.

 As for the annealing schedule of SantaQlaus, we employ two-fold stages corresponding to burn-in and refinement, though the update rules remain unchanged.
 We set the number of burn-in shots $s_{\mathrm{b}}$ such that the stage switches when that number of shots is used.
 From the start until $s_{\mathrm{b}}$ shots used, $\beta_t$ is given by $\beta_t = f_{0, s_{\mathrm{b}}}^{\beta_0, \beta_{\mathrm{b}}, a_1}(s_t)$ with parameters $\beta_0$, $\beta_{\mathrm{b}}$, and $a_1$ which determine the schedule, where $s_t$ is the total number of shots used before $t$-th iteration.
 We use $\beta_0 = 10$ throughout the simulations.
 Then, we use another schedule $f_{s_{\mathrm{b}}, s_{\max}}^{\beta_{\mathrm{b}}, \beta_{\mathrm{r}}, a_2}(s_t)$ for the refinement stage.
 In addition, we scale the inverse temperature as $\beta_t = f_{s_{\mathrm{b}}, s_{\max}}^{\beta_{\mathrm{b}}, \beta_{\mathrm{r}}, a_2}(s_t) / (r \eta_t)$ in the refinement stage, in order to avoid decrease in the number of shots when $\eta_t$ is decreased.
 We use $r=100$.
% 最大ゲインの評価：iCANSとの比較、オーバーヘッドに依存したスケーリング
% best と median いい方同士の比較で、到達速さ
%Finally, in order to demonstrate the performance of our optimizers in larger system sizes, we consider the
\begin{figure}
\centering
 \includegraphics[clip ,width=3.2in]{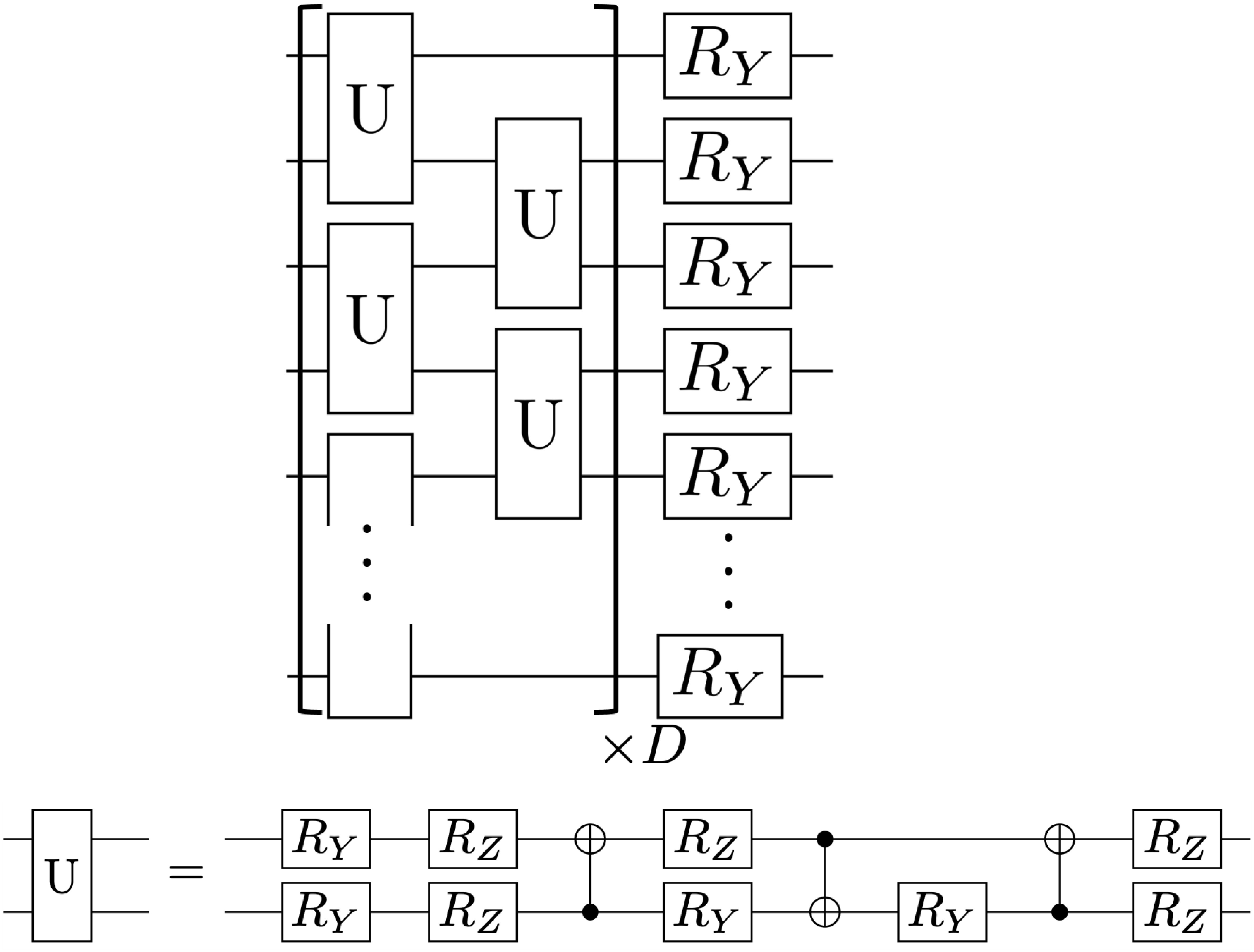}
 \caption{The parameterized quantum circuit used in the numerical simulation of the VQE task of a 1-dimensional transverse Ising spin chain.}
\label{fig-circuit-TFIM}
\end{figure}
\begin{figure*}
\centering
 \includegraphics[clip ,width=7.2in]{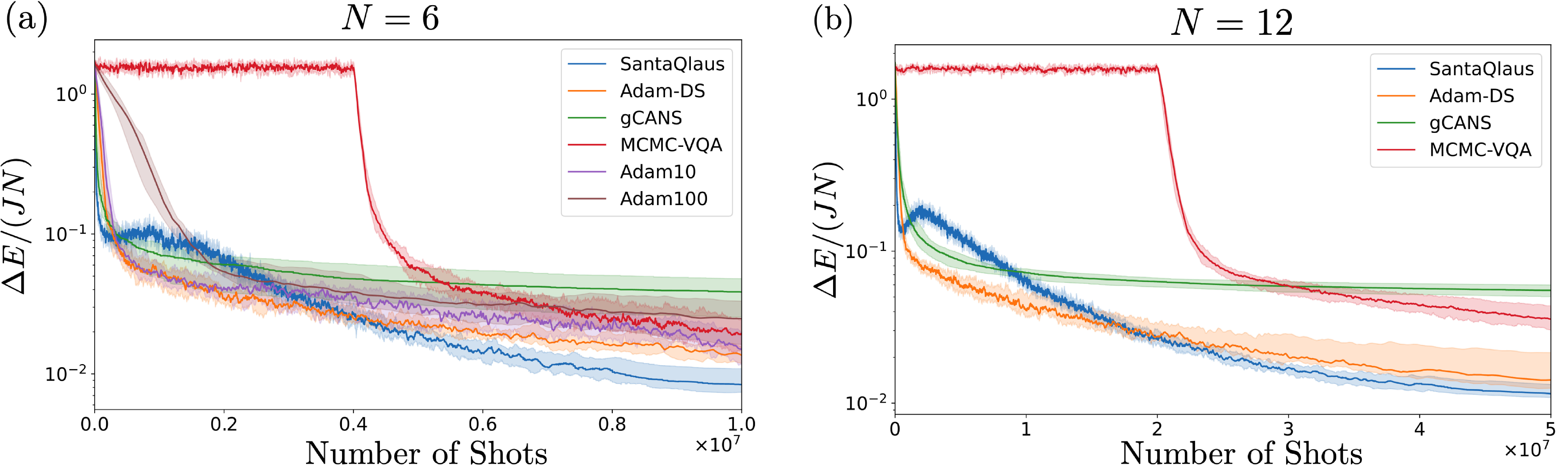}
 \caption{Comparison of performance of the optimizers for the VQE tasks of 1d transverse Ising spin chain with open boundary conditions. Every graph shows the median (solid curve) and the IQR (the highlighted region) of 20 runs of the optimizations with different random initial parameters. The exact expectation values are plotted to indicate the performance. (a) $N=4$. The per-site ground energy approximation precision $\Delta E / (J N)$ vs number of shots. (b) $N=12$. The value $\Delta E / (J N)$ vs number of shots.}
\label{fig-TFIM}
\end{figure*}

\subsection{VQE of 1-dimensional Transverse field Ising model}\label{ss_TFIM}

As a benchmark, we begin by a VQE task of the 1-dimensional transverse field Ising spin chain with open boundary conditions for the system size $N=6$ and 12, where the Hamiltonian is given as
\begin{align}
 H = -J\sum_{i=1}^{N-1}Z_i Z_{i+1} - g\sum_{i=1}^{N} X_i.
\end{align}
Our goal is to obtain the ground state energy by minimizing the expected energy as the loss function.
We consider the case with $g/J = 1.5$.
It is obvious that the interaction term $\sum_{i=1}^{N-1}Z_i Z_{i+1}$ or the transverse field term $\sum_{i=1}^{N} X_i$ can be measured at once.
We employ WDS to allocate the number of shots for them, as they form only two groups with similar weights.
If $s$ shots are used in total, we deterministically allocate $s J (N-1) / M$ to the former and $s g N / M$ to the latter, where $M = J(N-1)+g N$.
%The ratio of the number of shots used to measure the former to that for the latter is deterministically fixed as $J$ to $g$, as 
The results are shown in terms of the precision of the per-site energy $\Delta E / (J N)$, where $\Delta E$ denotes the difference of the exact energy expectation value evaluated at the parameters obtained by the optimization from the ground-state energy.
For this problem, we used the ansatz shown in Fig.~\ref{fig-circuit-TFIM} with $D=3$ following \cite{Kubler2020adaptiveoptimizer}.

%The results are shown in Fig.~\ref{fig-TFIM}.
%For both of cases with $N=6$ and $N=12$, SantaQlaus outperforms the other optimizers.
The results depicted in Fig.~\ref{fig-TFIM} shows the performance of various optimizers in the context of VQE tasks for 1d transverse Ising spin chains with open boundary conditions.
The median of the loss function and the interquartile range (IQR), i.~e.~the range between the first and third quartiles, are displayed.
Across scenarios with $N=6$ and $N=12$, SantaQlaus consistently demonstrates superior performance relative to the other optimizers.

For SantaQlaus, we employ $s_{\mathrm{b}}=0.8 s_{\max}$, $\beta_{\mathrm{b}}=\beta_{\mathrm{r}}=10^4$, $a_1=a_2=5$, and $a_{\mathrm{LR}}=0.5$ for both $N=6$ and $N=12$.
The learning rate exponent $a_{\mathrm{LR}}=0.1$ is used for Adam.
In Adam-DS, the number of shots is gradually increased from 4 to 100 (500) according to the function (\ref{stot_beta}) with $a=10$ for $N=6$ ($N=12$).
For MCMC-VQA, MCMC stage is chosen as $0.4 s_{\max}$, which is the best among $0.4 s_{\max}$, $0.6 s_{\max}$, and $0.8 s_{\max}$.
The learning rate exponent $a_{\mathrm{LR}}=0.3$ is used.

For $N=6$, we include Adam with fixed number of shots, denoted by Adam10 and Adam100 with each used number of shots.
As a result of the grid search, 10 is the best for the fixed number of shots among 10, 50, 100, 500, and 1000.
Adam-DS with tuned hyperparameters performs better than them, which implies a merit of simple dynamic shot-number increasing.
SantaQlaus is even faster than that and achieves better accuracy, indicating benefits beyond simple dynamic shot strategies.

For $N=12$, SantaQlaus attains the best median precision again.
While Adam-DS sometimes achieves comparable precision with SantaQlaus, it gets to poor local minima in some trials, as seen from its widespread of the IQR.
In stark contrast, SantaQlaus reliably maintains high precision, underscored by its narrower interquartile spread.

Despite its rapid initial progress, gCANS gets stuck in local minima at an early stage of optimization.
One contributing factor is that when the parameters enter a local mode or reach a stationary point, the shot-allocation rule of gCANS prescribes a large number of shots.
This is because the number of shots in gCANS is inversely proportional to the norm of the gradient.
MCMC-VQA does not seem to achieve sufficient mixing within this shot budget to exhibit the efficacy of MCMC sampling.

Overall, these results show that SantaQlaus is more efficient and effective in exploring the loss landscape compared to other optimizers, consistently finding better solutions.

\begin{figure*}
\centering
 \includegraphics[clip ,width=7.2in]{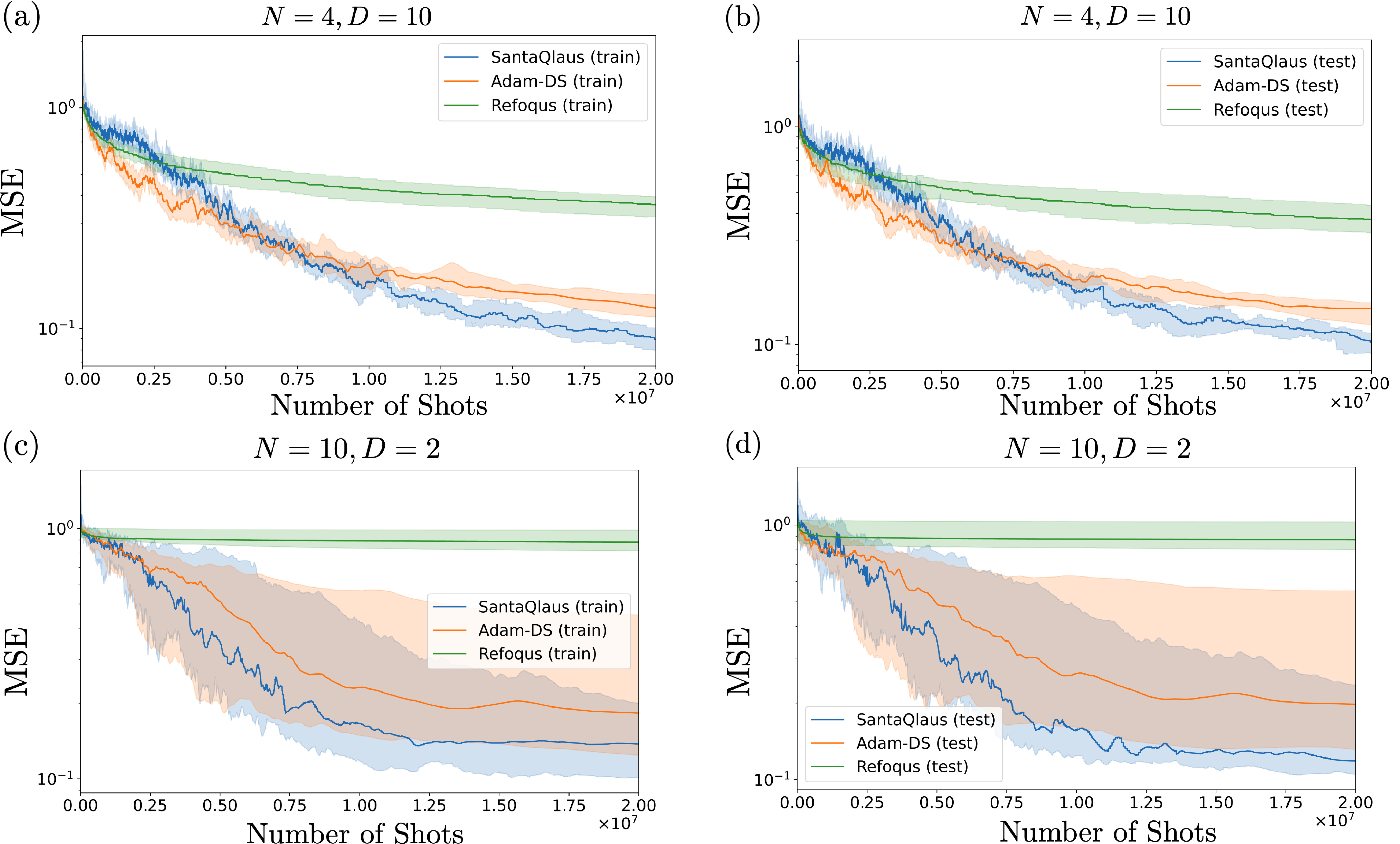}
 \caption{Comparison of performance of the optimizers for the regression task. Every graph shows the median (solid curve) and the IQR (the highlighted region) of 20 trials of the regression with different training data and the initial parameters. The MSE for the prediction obtained by the exact expectation values is plotted to indicate the performance. (a) 4-qubit 10 layers. MSE for train data vs number of shots. (b) 4-qubit 10 layers. MSE for test data vs number of shots. (c) 10-qubit 2 layers. MSE for train data vs number of shots. (d) 10-qubit 2 layers. MSE for test data vs number of shots.}
\label{fig-regress}
\end{figure*}

\subsection{Benchmark regression task}\label{ss_regression}

We next test SantaQlaus on a QML regression task investigated in Ref.~\cite{Jerbi:2023aa}.
A QNN used here consists of feature encoding unitary $U(\bm{x})$ and the trainable unitary $V(\bm{\theta})$ which form the model state $\ket{\psi(\bm{x};\bm{\theta})} = V(\bm{\theta})U(\bm{x})\ket{0}$.
Given some set $\mathcal{D} = \{\bm{x}_1,\cdots, \bm{x}_N\}$ of input data vectors, the label $g(\bm{x})$ of each data $\bm{x}$ is generated by the model with a randomly chosen target parameter $\bm{\theta}^* \in [0, 2\pi)$ as
\begin{align}
 g(\bm{x}) = w^{*} \bra{\psi(\bm{x};\bm{\theta}^*)} Z_1 \ket{\psi(\bm{x};\bm{\theta}^*)},
\end{align}
where $w^*$ is a scale factor that sets the standard deviation of the labels to $1$ over the dataset, and $Z_1$ denotes the Pauli $Z$ on the first qubit.
Our goal is to do a regression of these labels by $w \bra{\psi(\bm{x};\bm{\theta}^*)} Z_1 \ket{\psi(\bm{x};\bm{\theta}^*)}$ via tuning $w$ and $\bm{\theta}$ without knowing the target parameters.
We use the MSE as the loss function.
As the correct parameters exist, the representability of the ansatz does not matter in this task.

As the input data, we use a dimensionality-reduced feature vectors of fashion MNIST \cite{Xiao:2017aa} via principal component analysis following Ref.~\cite{Jerbi:2023aa}.
We use $M=1100$ data points as the whole dataset.
In each training, the size of the training dataset is $880$ and the test data size is $220$.
We use the same feature encoding circuit $U(\bm{x})$ and trainable circuit $V(\bm{\theta})$ as Ref.~\cite{Jerbi:2023aa}.
The feature encoding is similar to the one proposed in Ref.~\cite{Havlicek:2019ab}, which is given as:
\begin{equation}
U(\bm{x}) \ket{0^{\otimes N}} = U_z(\bm{x}) H^{\otimes N} U_z(\bm{x}) H^{\otimes N} |0^{\otimes N}\rangle
\end{equation}
with
\begin{equation}
 U_z(\bm{x}) = \exp \left( -i\pi \left[ \sum_{i=1}^{N} x_i Z_i + \sum_{j=1, j>i}^{N} x_i x_j Z_i Z_j \right] \right)
\end{equation}
for $N$-qubit system, where $H$ denotes the Hadamard gate.
As for the trainable circuit $V(\theta)$, these are composed of $D$ layers of single-qubit rotations $R(\theta_{i,j})$ on each of the qubits, interlaid with $CZ = |1\rangle\langle 1| \otimes Z$ gates between nearest neighbours in the circuit, where $R(\theta_{i,j}) = R_X(\theta_{i,j,0}) R_Y(\theta_{i,j,1}) R_Z(\theta_{i,j,2})$.
%We choose the number of layers $ L $ as a function of the number of qubits $ n $ in the circuit, such that the number of parameters $ 3nL $ is approximately 90 at all system sizes. Specifically, from $ n = 2 $ to 12, we have $ L = 15,10,7,6,5,4,4,3,3,3,3 $ respectively.
We implemented simulations for the case of $N=4$, $D=10$ (133 parameters), and $N=10$, $D=2$ (91 parameters).
We employ $s_{\mathrm{b}}=0.5 s_{\max}$, $\beta_{\mathrm{b}}=5000 (500)$, $\beta_{\mathrm{r}}=10^4 (10^3)$, and $a_1=a_2=3$ for $N=4 (N=10)$.
The learning rate exponent $a_{\mathrm{LR}}=0.3$ is used for both SantaQlaus and Adam-DS.
In Adam-DS, the number of shots is gradually increased from 4 to 100 according to the function (\ref{stot_beta}) with $a=2$.
For both optimizers, the batch size is gradually increased from 2 to 16 according to the function (\ref{stot_beta}) with $a=2$.

The performance of different optimizers for the regression task is shown in Fig.~\ref{fig-regress}.
Clearly, SantaQlaus demonstrates superior performance compared to other optimizers, achieving the lowest median MSE for both training and test data.
Regarding the 4-qubit case shown in (a) and (b) of Fig.~\ref{fig-regress}, while Adam-DS displays some optimization progression, SantaQlaus consistently achieves lower MSE, indicating its enhanced accuracy and efficient use of shot resources.
%The 10-qubit task appears more challenging, as the learning curves suggests the presence of numerous poor local minima and plateaus.
Turning to the 10-qubit scenario in graphs (c) and (d) of Fig.~\ref{fig-regress}, the optimization landscape appears notably challenging.
Here, Refoqus is unable to escape from a plateau without showing significant improvement as the number of shots increases.
Adam-DS, while presenting some improvement in the median MSE, has a large upper quartile, suggesting that it often becomes trapped in less optimal regions of the optimization landscape.
%Although Adam presents some improvement in the median MSE, its large upper quartile indicates that it is frequently trapped to poor local minima or plateaus.
In contrast, SantaQlaus consistently delivers a lower median MSE and exhibits a decreasing upper quartile, highlighting its robust capability in navigating and optimizing even in such challenging landscapes.
%In contrast, SantaQlaus not only outperforms Adam in median MSE but also displays a consistent decline in the upper quartile MSE, emphasizing its adaptability and effectiveness even in such intricate optimization landscapes.

\begin{figure*}
\centering
 \includegraphics[clip ,width=7.2in]{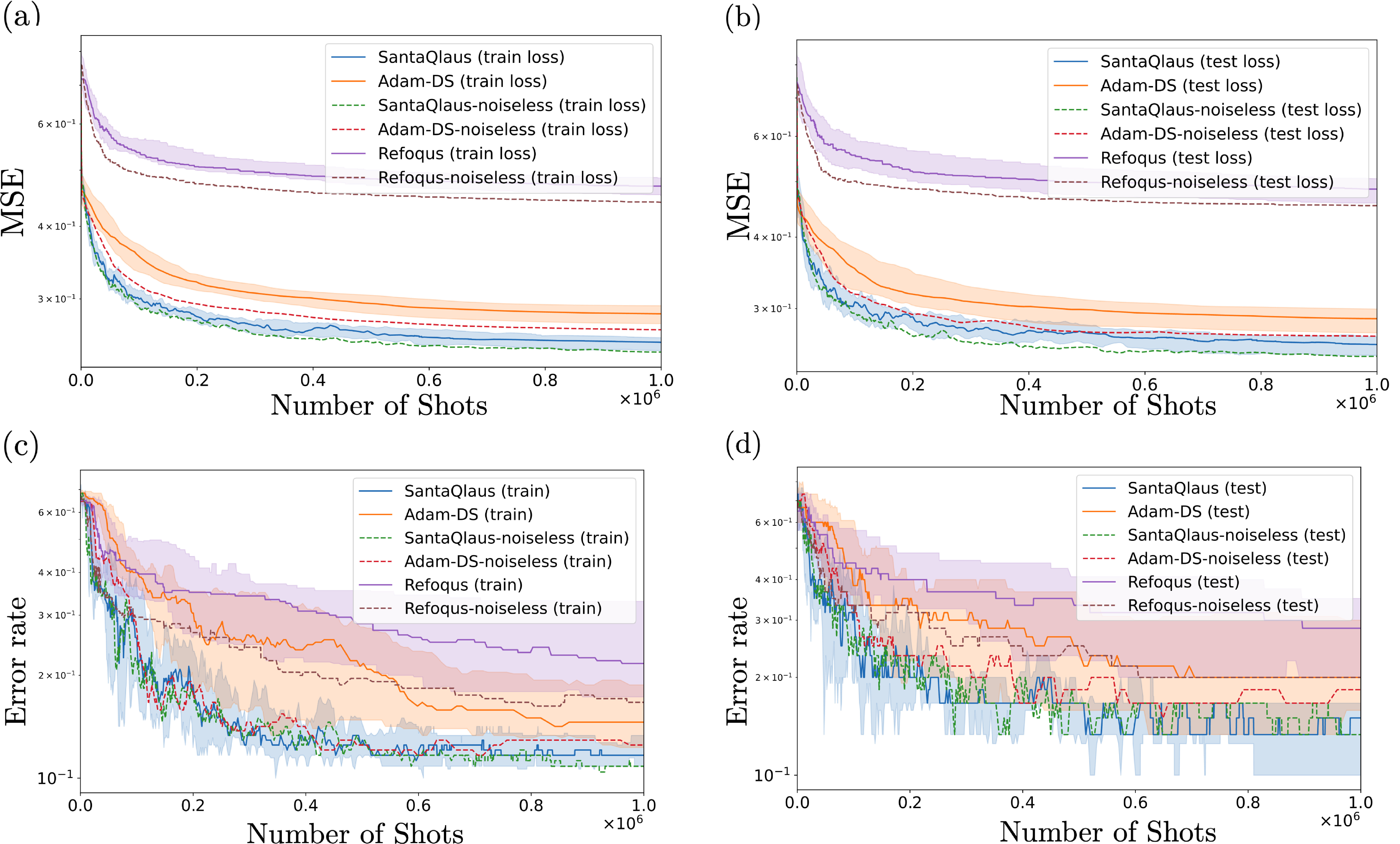}
 \caption{Comparison of performance of the optimizers for the classification of the Iris dataset. Every graph shows the median (solid curve) and the IQR (the highlighted region) of 20 trials of the training with different training data and the initial parameters. The dashed curves depict the median of learning curves in the absence of the depolarizing noise (only shot noise). Plotted MSE and error rate are computed from the exact expectation values without hardware noise, while training are done with finite number of shots under the influence of the local depolarizing noise. (a) MSE for train data vs number of shots. (b) MSE for test data vs number of shots. (c) Error rate for train data vs number of shots. (d) Error rate for test data vs number of shots.}
\label{fig-Iris}
\end{figure*}

\subsection{Classification of Iris dataset with local depolarizing noise}\label{ss_iris}
Finally, we test SantaQlaus on a classification task using the Iris dataset \cite{Anderson:1936aa}.
The Iris dataset comprises three classes of Iris flowers, with 50 samples per class, and includes four features for each sample.
We use the normalized input data $\bm{z}$ given by $z_{i,j} = (x_{i,j} - \min_k (x_{k,j}))/(\max_k(x_{k,j}) - \min_k(x_{k,j}))$, where $x_{i,j}$ denotes the $j$-th feature of the $i$-th data point.
We randomly select 120 data points used for the training.
Then, the remaining 30 data points are used as the test data.

We employ the ansatz $\ket{\psi(\bm{x}; \bm{\theta})}$ given by the same feature encoding and trainable circuits as those for the previous QML benchmark in Sec.~\ref{ss_regression} with $N=4$ and $D=4$.
We use $Z_1\otimes Z_2$ as the observable to be measured, which yields two bits $(b_1, b_2)$ as an outcome.
Then, we assign a label of the class $y(b_1,b_2):= b_1 + 2 b_2 \;(\mathrm{mod} 3)$ to each outcome.
The label prediction for each data point $\bm{x}$ is determined by the most frequently occurring value of $y(b_1,b_2)$, based on repeated measurements of $\ket{\psi(\bm{x}; \bm{\theta})}$.
Hence, the label value with the highest probability emerges as the predicted label with infinitely many measurements.
As the loss function to be minimized, we use the MSE of the success probability given as
\begin{align}
 L(\bm{\theta}) = \frac{1}{M}\sum_{i=1}^M (1 - p(\bm{x}_i, \bm{\theta}))^2,
\end{align}
where $p(\bm{x}_i, \bm{\theta})$ denotes the probability of obtaining the correct label for the data point $\bm{x}_i$ with the model parameter $\bm{\theta}$.
This is the same as the mean squared failure probability.
Because this is MSE, we obtain unbiased estimators of the gradient and its variance as we have seen in Sec.~\ref{ss_loss} and \ref{ss_normal}.
That is the reason of this choice of the loss function in this benchmark.

In this simulation, hardware noise is incorporated, specifically modeled as local depolarizing noise. To represent this, a single-qubit (two-qubit) depolarizing channel is introduced after every single-qubit (two-qubit) gate, with the exception of $Z$-rotation gates. The error probabilities assigned are $10^{-3}$ for single-qubit gates and $10^{-2}$ for two-qubit gates.

As for the hyperparameters specific to this benchmark, we employ $s_{\mathrm{b}}=0.5 s_{\max}$, $\beta_{\mathrm{b}}=10^4$, $\beta_{\mathrm{r}}=5\times 10^4$, and $a_1=a_2=3$ for SantaQlaus.
The learning rate exponent $a_{\mathrm{LR}}=0.3$ is used for both SantaQlaus and Adam-DS.
In Adam-DS, the number of shots is gradually increased from 4 to 10 according to the function (\ref{stot_beta}) with $a=3$.
For both optimizers, the batch size is gradually increased from 2 to 16 according to the function (\ref{stot_beta}) with $a=1$.

The performance of different optimizers for this classification task is shown in Fig.~\ref{fig-Iris}.
Here, the error rate is defined as the proportion of incorrectly predicted labels to the total number of data points.
Both MSE and the error rate are calculated by the exact expectation values in the absence of noise, to evaluate the achievable performance of the obtained model.
To account for misclassification arising from statistical errors, we assume a worst-case scenario.
In this scenario, a data point $\bm{x}_i$ is considered misclassified if the difference $\Delta p(\bm{x}_i)$ between the highest and the second-highest label probabilities is less than $2\epsilon$, where $\epsilon$ is a positive constant.
This approach is relevant because if the statistical error in probability distribution estimation exceeds $\epsilon$ and $\Delta p(\bm{x}_i)$ is smaller than $2\epsilon$, there is a risk of predicting an incorrect label, even when the highest probability corresponds to the correct label.
To estimate the probability within the precision of $\epsilon$, we need $1/\epsilon^2$ times sampling overhead for both the direct estimation using a quantum device, and the acquisition of a classical surrogate \cite{Schreiber:2023aa}.
Here, we choose $\epsilon = 10^{-2}$ corresponding to the sampling overhead $10^4$.

According to the resulting learning curves in Fig.~\ref{fig-Iris}, SantaQlaus demonstrates the highest performance even in the presence of the local depolarizing noise.
For both the MSE and the error rate, we can clearly see that the ability of SantaQlaus to efficiently exploring better optimal parameters surpasses the others.
Notably, our results imply that SantaQlaus is much more robust to the hardware noise than the other methods.
Indeed, the presence of hardware noise significantly impairs the learning performance of the Adam-DS optimizer.
In contrast, SantaQlaus maintains nearly consistent performance levels under the same noise.
This robustness may be attributed to the efficient exploration strategy of SantaQlaus, which leverages QSN with its adaptively adjusted variance.
It is suggested that this strategy is also effective in navigating challenging landscape which may be caused by hardware noise.

\section{Conclusion}
\label{jzbmlzhd}
In this study, we introduced SantaQlaus, an optimizer designed to strategically leverage inherent QSN for efficient loss landscape exploration while minimizing the number of quantum measurement shots.
The algorithm is applicable to a broad spectrum of loss functions encountered in VQAs and QML, including those that are non-linear.
Incorporating principles from the classical Santa algorithm \cite{Chen:2016aa}, SantaQlaus exploits annealed QSN to effectively evade saddle points and poor local minima.
The algorithm adjusts the number of measurement shots to emulate appropriate thermal noise based on the asymptotic normality of QSN in the gradient estimator.
This adjustment requires only a small classical computational overhead for variance estimation.
Moreover, the update rule of our algorithm includes thermostats from Santa that provide robustness against estimation errors of the variance of QSN.
SantaQlaus naturally attains resource efficiency by initiating the optimization process with a low shot count during the high-temperature early stages, and gradually increasing the shot count for more precise gradient estimation as the temperature decreases.

We have demonstrated the efficacy of SantaQlaus through numerical simulations on benchmark tasks in VQE, regression task, and a multiclass classification under the influence of the local depolarizing noise.
Our optimizer consistently outperforms existing algorithms, which often get stuck in suboptimal local minima or flat regions of the loss landscape.
Our results imply that compared to shot-adaptive strategies like gCANS, SantaQlaus excels in directly addressing the challenges in the loss landscape rather than merely maximizing iteration gains.
SantaQlaus also exhibits advantages over basic shot-number annealing approaches like Adam-DS.
Moreover, our simulation implies that SantaQlaus is robust against hardware noise, which may also highlight the efficiency of the exploration strategy of SantaQlaus leveraging QSN.

Looking ahead, additional research is needed to assess performance of SantaQlaus in experiments.
The demonstrated robustness of SantaQlaus against hardware noise indicates its potential for promising results in practical experiments.
This aspect warrants comprehensive investigation to fully understand and leverage its capabilities in real-world applications.
%As we highlighted in our simulation, we can expect that SantaQlaus has a robustness against hardware noise.
Incorporating QEM techniques into SantaQlaus also offers a promising route for experimental deployments.
For QNNs, combining SantaQlaus with a recent noise-aware training strategy \cite{Wang:2022aa}  could potentially enhance robustness and efficiency under realistic conditions.
Incorporating advanced preconditioning techniques, such as Fisher information, may provide further improvements. These avenues remain open for future exploration.
%It would lead to further improvement to incorporate more general preconditioner such as the Fisher information into SantaQlaus.
%We leave them for a future work.
%%%%%%%%%%%%%%%%%%%%%%%%%%%%%%%%%%%%%%%%%%%%%%%%%%%%%%%%%%%%%%%%%%%%%%%
\begin{acknowledgments}
This work is supported by MEXT Quantum Leap Flagship Program (MEXT QLEAP) Grant Number JPMXS0120319794, and JST COI-NEXT Grant Number JPMJPF2014.
\end{acknowledgments}

\bibliographystyle{apsrev4-1}
\bibliography{papers}

\end{document}